\documentclass[12pt, a4paper]{article}
\usepackage[textwidth = 7in, textheight=25cm,nohead]{geometry}
\usepackage{ulem}
\usepackage{graphicx, amssymb, amsmath, amsthm, amstext, booktabs, float, multirow, subfigure, rotating,natbib}
\usepackage{setspace}
\numberwithin{equation}{section} \numberwithin{figure}{section} \numberwithin{table}{section}

\bibpunct{(}{)}{,}{a}{,}{,}

\def\ppn{\vskip 6pt \noindent }
\def\R{{\mathbb{R}}}

\def\E{{\mathbb{E}}}
\newcommand{{\Xs}}{{\cal X}}
\newcommand{{\Ls}}{{\cal L}}
\newcommand{{\Ss}}{{\cal S}}
\newcommand{{\Gs}}{{\cal G}}
\newcommand{{\Hs}}{{\cal H}}
\newcommand{{\Ns}}{{\cal N}}
\newcommand{{\Is}}{{\cal I}}
\newcommand{{\Bs}}{{\cal B}}
\newcommand{{\Cs}}{{\cal C}}
\newcommand{{\Rs}}{{\cal R}}
\newcommand{{\pp}}{{\mathbf p}}
\newcommand{{\KK}}{{\mathbf K}}
\newcommand{{\HH}}{{\mathbf H}}
\newcommand{{\II}}{{\mathbf I}}
\newcommand{{\yy}}{{\mathbf y}}
\newcommand{{\dou}}{$\leadsto$\ }

\newcommand{{\phphx}}{{\phi(\Phi^{-1}(x))}}
\newcommand{{\phphi}}{{\phi(\Phi^{-1}(X_i))}}
\newcommand{{\Phx}}{{\Phi^{-1}(x)}}
\newcommand{{\PhXi}}{{\Phi^{-1}(X_i)}}
\newcommand{{\fC}}{{\hat{f}_{X\text{,C}_2}}}
\newcommand{{\fGC}}{{\hat{f}_{X\text{,GC}}}}
\newcommand{{\fdiff}}{{\hat{f}_{X\text{,diff}}}}
\newcommand{{\fhatTX}}{{\hat{f}^{(T)}_X}}
\newcommand{{\fhatTXstar}}{{\hat{f}^{*(T)}_X}}

\DeclareMathOperator{\var}{\mathbb{V}ar}

\begin{document}
\title{Probit transformation for kernel density estimation on the unit interval}
\author{\sc{Gery Geenens}\thanks{Corresponding author: ggeenens@unsw.edu.au, School of Mathematics and Statistics, University of New South Wales, Sydney, NSW 2052 (Australia), tel +61 2 938 57032, fax +61 2 9385 7123 }\\School of Mathematics and Statistics,\\ University of New South Wales, Sydney, Australia}
\date{}
\maketitle
\thispagestyle{empty} 

\begin{abstract}
Kernel estimation of a probability density function supported on the unit interval has proved difficult, because of the well known boundary bias issues a conventional kernel density estimator would necessarily face in this situation. Transforming the variable of interest into a variable whose density has unconstrained support, estimating that density, and obtaining an estimate of the density of the original variable through back-transformation, seems a natural idea to easily get rid of the boundary problems. In practice, however, a simple and efficient implementation of this methodology is far from immediate, and the few attempts found in the literature have been reported not to perform well. In this paper, the main reasons for this failure are identified and an easy way to correct them is suggested. It turns out that combining the transformation idea with local likelihood density estimation produces viable density estimators, mostly free from boundary issues. Their asymptotic properties are derived, and a 
practical cross-validation bandwidth selection rule is devised. Extensive simulations demonstrate the excellent performance of these estimators compared to their main competitors for a wide range of density shapes. In fact, they turn out to be the best choice overall. Finally, they are used to successfully estimate a density of non-standard shape supported on $[0,1]$ from a small-size real data sample.

\medskip

\noindent \textbf{Keywords:} transformation kernel density estimator; boundary bias; local likelihood density estimation; local log-polynomial density estimation.
\end{abstract}

\section{Introduction} \label{sec:intro}

Kernel density estimation is a standard nonparametric method for estimating a probability density without making any rigid assumption about the distribution of the data, see for instance \cite{Wand95} and \cite{Simonoff96}. Given a sample $\{X_1,X_2,\ldots,X_n\}$ of i.i.d.\ observations from a univariate distribution $F_X$ admitting a density $f_X$, the conventional kernel density estimator is
\begin{equation} \hat{f}_X(x) = \frac{1}{nh}\sum_{i=1}^n K\left(\frac{x-X_i}{h} \right). \label{eqn:baskde} \end{equation}
The `kernel' function $K$ is usually taken to be a symmetric unimodal probability density and the `bandwidth' $h$ is the parameter that controls the smoothness of the final estimate. 
The huge amount of literature on kernel density estimation testifies that, provided that the value of $h$ is sensibly selected, estimator (\ref{eqn:baskde}) is a reliable one for estimating $f_X$ in a flexible way.

\ppn It is, however, well known that it is not appropriate when the support of $f_X$ is bounded. The reason is that the estimator does not `feel' the support boundaries and, for $x$ close to a boundary, places some positive mass outside that support. This results in a significant bias which may prevent the estimator from being consistent in those areas. As support restrictiveness is common in practice, for instance when $X$ is known to be positive, curing that boundary bias problem has attracted a lot of attention in the literature, see, {\it inter alia}, \cite{Schuster85,Muller91,Lejeune92,Jones93,Jones96,Cowling96,Cheng97,Zhang98,Zhang00,Zhang99,Chen00,Hall02,Park03,Scaillet04} and \cite{Karunamuni05}. Recently, \cite{Dai10} suggested a simple correction based on utilizing a local bandwidth close to the boundary. Of main interest in this paper, though, will be a procedure close in spirit to that suggested in \cite{Marron94}, namely transforming the variable of interest into another one whose density 
estimation should be free from boundary problems, and transform that estimate back into the initial scale. Although this transformation method seems very natural and has been around for a long time, a simple and efficient practical implementation of this methodology is yet to be developed, to the best of this author's knowledge. \cite{Marron94}'s method, for instance, has often been labeled `very complicated' in the subsequent literature.

\ppn This paper actually investigates the transformation method in the case where the support of $f_X$ is compact, for instance the unit interval $\Is=[0,1]$. As there is no loss of generality in considering the $[0,1]$-support case only, for any compact domain can be mapped onto $[0,1]$ by straightforward linear rescaling, this case will be the only focus here.
When the density is supported on $[0,1]$, it is clear that the above mentioned bias issues are even more disturbing. Indeed, in that case both boundaries are likely to affect the performance of the density estimator over a major part of the support of $f_X$. Of course, each boundary can be considered separately and some {\it ad hoc} surgery on the estimator can be performed close to them using methods like those cited in the previous paragraph. It seems, however, more natural to devise estimators which automatically take the constrained nature of $X$ into account from the outset. In fact, such estimators were indeed suggested in \cite{Chen99}, \cite{Jones07a,Jones07b} and recently, in a wider framework, \cite{Botev10}. 

\ppn The key idea is that, in order to properly handle the bounded support $\Is$ of $f_X$, a kernel estimator like (\ref{eqn:baskde}) should use a kernel function $K$ which is also supported on $\Is$, and this for all combinations of $x$, $X_i$ and $h$. This would indeed prevent it from assigning positive weight outside $\Is$ and the consequences thereof. Accordingly, \cite{Chen99} suggested to take $K$ to be a suitably parameterized Beta density, defining the {\it Beta kernel estimator} $\fC$. On the other hand, \cite{Jones07a,Jones07b} rather suggested to use as kernels the conditional densities extracted from a bivariate Gaussian Copula, leading to their {\it Gaussian Copula kernel estimator}, $\fGC$. The notations $\fC$ and $\fGC$ have been borrowed from \cite{Jones07a,Jones07b} (except for the subscript `$X$').

\ppn The asymptotic bias and variance of the Beta kernel estimator were given in \cite{Chen99}, and further asymptotic properties were studied in \cite{Bouezmarni03} and \cite{Zhang10}. As beta-kernel estimators are directly related to Bernstein polynomial smoothing, relevant information can also be found in \cite{Brown99,Bouezmarni07} and \cite{Leblanc12}. \cite{Jones07a,Jones07b} derived similar theoretical properties for their Gaussian Copula estimator and provided a detailed comparison between $\fC$ and $\fGC$ both in theory and in practice. They stressed the overall very good practical performance of the Beta kernel estimator, but they also pointed out some shortcomings that $\fGC$ was aimed at correcting. In particular, $\fC$ has, by construction, a propensity for estimating $f(0)$ by a nonzero finite value in all cases. Hence it struggles to appropriately estimate densities such that $f(0)=0$ (same at $x=1$) or densities admitting a pole at one (or both) of its boundaries. Finally, \cite{Jones07a,
Jones07b} suggested a `rule-of-thumb' reference bandwidth selector for easily choosing the smoothing parameter for both the Beta and the Gaussian Copula kernel estimators. Their comprehensive simulation studies showed that $\fC$ and $\fGC$ were roughly level on overall performance for a wide range of density shapes on $[0,1]$. \cite{Botev10}'s idea is somewhat different and has to be replaced within the frame of a kernel density estimation approach elegantly based on the properties of linear diffusion processes. This estimator, called the {\it diffusion estimator} and denoted $\fdiff$ in this paper, was claimed to cure boundary bias problems as well. Developing and investigating a competitor for these three estimators is actually the main aim of this paper.

\ppn As stated earlier, the estimator proposed here is based on transformation. Combining kernel density estimation and transformation is not a novel idea, though. This was first suggested in \citet[Chapter 9]{Devroye85} and \citet[Sections 2.9 and 2.10]{Silverman86}, and then studied in depth in \cite{Wand91,Park92,Ruppert94,Hossjer95} and \cite{Yang99}. Later, it was taken up in \cite{Bolanceal08,Buchlarsen05} and \cite{Markovich05}. There, the transformation did not aim at taking care of boundary bias, but rather dealing with other density features known to make the estimation difficult, such as skewness, smoothness inhomogeneity or heavy tails. Some exceptions, explicitly targeting boundary bias reduction, are the above mentioned \cite{Marron94}, \cite{Karunamuni06} and \cite{Koekemoer08}. In fact, all those papers attempt to obtain, via transformation of the initial data, a pseudo-sample coming from a distribution `easy to estimate' in some sense, like the Uniform or the Normal distribution. Of course, 
the transformation able to produce the selected target distribution depends on the initial distribution of the original data, and should therefore be estimated from them. This need for estimating a tailored transformation unfortunately makes those procedures somewhat unwieldy and in a sense spoils the simplicity of the initial transformation idea.

\ppn What is suggested here markedly departs from those previous contributions in that a fixed transformation is considered, without any attempt to produce any particular distribution. This can be motivated inversely to what has been said for the Beta and Gaussian Copula kernel estimators. Instead of working with unusual kernels whose supports are always the support $\Is$ of $X$, here that is $\Is$ which is transformed such that it always matches the support of usual kernels for any combination of $x$, $X_i$ and $h$, that is the whole $\R$. Loosely speaking, here the transformation just aims at sending away the boundaries to $\pm \infty$. This can potentially be achieved by any transformation $T: [0,1] \to \R$, continuous and increasing, such that $\lim_{x\to 0} T(x) = -\infty$ and $\lim_{x\to 1} T(x) = +\infty$. A natural choice seems to be the standard normal quantile function, denoted $\Phi^{-1}$ throughout the paper, hence the name of {\it probit-transformation} for the kernel estimators studied in this 
paper. 

\ppn It is acknowledged that this idea was mentioned in \citet[Section 6.2]{Jones07b}, but those authors reported (p.\ 17) that ``(...) {\it standard kernel density estimator with rule-of-thumb bandwidth applied to probit-transformed versions of the data} (...) {\it did not compete well }(...)'' against \cite{Chen99}'s and \cite{Jones07a}'s estimators. This was also pointed out in other papers implementing the idea as-is. Yet, upon reflection, it seems clear that standard kernel estimation using a standard bandwidth is certainly not the right choice in this very situation, as discussed at length in the next section. In this paper, a totally viable probit-transformation kernel density estimator is devised. The good point is that, the transformation being fixed, the suggested procedure is barely more complicated than any other conventional nonparametric density estimation.

\ppn In the next section, the idea of probit transformation is introduced in more detail, the reasons why a straightforward application of it is not expected to give acceptable results are set out, and a way to enhance that naive estimator is suggested. Section \ref{sec:asymptotics} studies the theoretical asymptotic properties of the suggested estimators. 
Section \ref{sec:bandwidth} studies the always crucial point of bandwidth selection in this particular setting. Some practical rules are devised, of which some will prove best in practice in the simulation study of Section \ref{sec:sim}, where the different estimators discussed in the paper are put to the test. In Section \ref{sec:realdata}, the density of a real data sample is successfully estimated via the proposed methodology. Finally, Section \ref{sec:ccl} concludes and suggests some way to continue this research.

\section{Probit-transformation kernel density estimation}\label{sec:transf}

\subsection{Probit-transformation}

As anticipated in the previous section, the suggested procedure for estimating a density supported on the unit interval is the following. Denote $S = \Phi^{-1}(X)$ the variable of interest in the probit-transformed domain. It is clear that, if $f_X(x)>0$ almost everywhere on $[0,1]$ (which will be assumed throughout this paper), then $S$ has unbounded support. From standard arguments, the density of $S$, say $f_S$, is related to that of $X$ by $f_S(s) = f_X(\Phi(s))\phi(s)$
for all $s \in \R$, where $\Phi$ is the standard normal cumulative distribution function, and $\phi$ its density. Although obvious, it is stressed that $S$ has no particular reason for being normally distributed. In fact, that will only be the case if the density of $X$ on $[0,1]$ is one of the conditional densities extracted from a bivariate Gaussian copula, including as particular case $X \sim U_{[0,1]}$ (then $S \sim \Ns(0,1)$). It is on this fact that \cite{Jones07a,Jones07b}'s `Rule-of-Thumb' bandwidth selector for $\fGC$ is based.

\ppn Naturally, one can also write $f_X(x) = f_S(\Phi^{-1}(x))/\phi(\Phi^{-1}(x))$ for all $x \in [0,1]$ (to be understood as $\lim_{s \to \pm \infty} f_S(s)/\phi(s)$ at $x=0/1$). It therefore follows that any estimator $\hat{f}_S$ of $f_S$ instantly provides an estimator of $f_X$, viz.
\begin{equation} \hat{f}^{(T)}_X(x) = \frac{\hat{f}_S(\Phi^{-1}(x))}{\phi(\Phi^{-1}(x))}, \label{eqn:ratiodens}
 \end{equation}
where the superscript $(T)$ refers to the idea of transformation. As $S$ is unconstrained, $\hat{f}_S$ is free from boundary problems, and mostly so should be $\hat{f}^{(T)}_X$. It is also clear that $\hat{f}^{(T)}_X(x)$ cannot allocate any positive probability outside the genuine support $\Is=[0,1]$ of $f_X$.

\subsection{The naive estimator}

\ppn From the `pseudo'-sample $\{S_1,\ldots,S_n\}$, with $S_i = \Phi^{-1}(X_i)$, $i=1,\ldots,n$, a first idea is to estimate $f_S$ with a standard kernel density estimator on the real line, that is an estimator like (\ref{eqn:baskde}): 
\begin{equation} \hat{f}_S(s) = \frac{1}{nh} \sum_{i=1}^n K\left(\frac{s-S_i}{h} \right) \label{eqn:naivedensS} \end{equation}
 Back-transforming this to the $X$-domain through (\ref{eqn:ratiodens}) directly yields
\begin{equation} \hat{f}^{(T)}_X(x) = \frac{1}{nh \phphx} \sum_{i=1}^n K\left(\frac{\Phx-\PhXi}{h} \right).  \label{eqn:naivetransf} \end{equation}
Noting that $(\Phx)'=1/\phphx$, one can write the linear Taylor expansion $\PhXi \approx \Phx + (X_i-x)/\phphx$ for $X_i$ close to $x$, whence
\begin{equation*} \hat{f}^{(T)}_X(x) \approx \frac{1}{nh \phphx} \sum_{i=1}^n K\left(\frac{x-X_i}{h \phphx} \right). \label{eqn:naivetransflocal} \end{equation*}
Thus, $\hat{f}^{(T)}_X(x)$ essentially behaves like a conventional kernel estimator with a local bandwidth $h^*(x) = h\phphx$, i.e.\ a bandwidth depending on the estimation point $x$, as already stressed in \cite{Wand91}. 
As $\lim_{x \to 0/1} \phphx = 0$, this shows that $\hat{f}^{(T)}_X(x)$ intrinsically uses smaller and smaller bandwidths when $x$ is approaching 0 or 1, and this is essentially how it attempts to cure the boundary bias problem. As such, this is similar to what \cite{Dai10} suggested. 

\begin{figure}[h]
\centering
\includegraphics[scale=0.3]{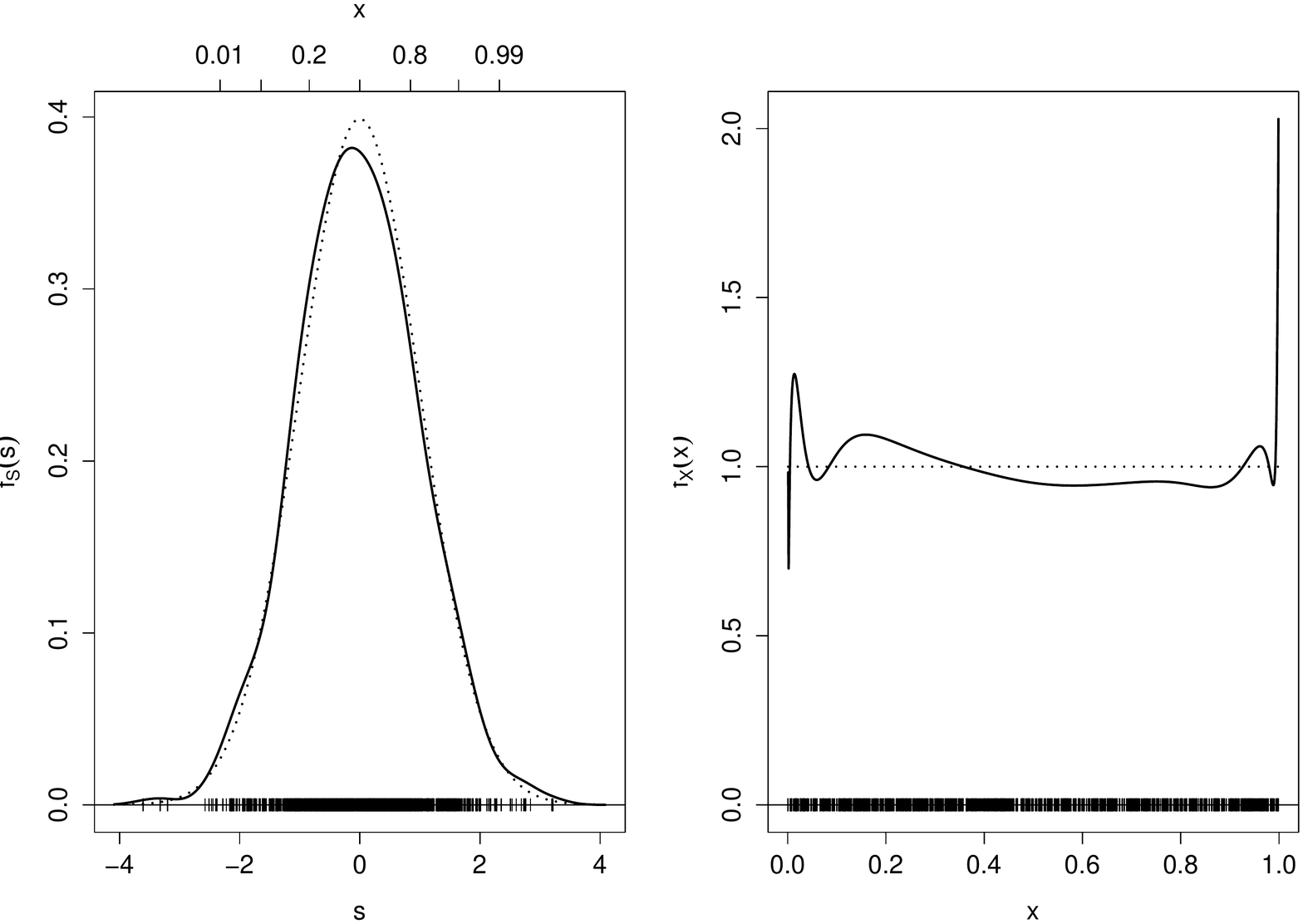}
\caption{True (dotted) and estimated (plain) densities in the S-domain and in the X-domain for a simulated sample ($n=1000$) from $U_{[0,1]}$. The kernel was the standard normal density and the bandwidth selected by Normal reference rule was $h=0.27$. Corresponding $X$-scaling is shown on top of the left panel for benchmark.}
\label{fig:naivedens}
\end{figure}

\ppn Below, this estimator (\ref{eqn:naivetransf}) will be called the {\it naive probit-transformation kernel density estimator}, as a simple example suffices to illustrate the problems that it faces. A sample of size $n=1000$ was generated from the $U_{[0,1]}$- distribution, and its density was estimated by (\ref{eqn:naivetransf}). Here, $f_S$ is the standard normal density, therefore in (\ref{eqn:naivedensS}) $K$ was chosen to be the Gaussian kernel and $h$ was selected by Normal reference rule \citep[Section 3.2.1]{Wand95}, hence close to be optimal. Figure \ref{fig:naivedens} shows the estimated densities in both the $S$-domain (left) and the $X$-domain (right). 

\ppn In the left-panel, (\ref{eqn:naivedensS}) does reasonably good at estimating the normal density in the $S$-domain. When this $\hat{f}_S$ is transformed back to the $X$-domain, however, it produces the poor estimate in the right panel. In the interior, it decently estimates the Uniform density. However, towards 0 and 1, where the attention mostly lies ($ \hat{f}^{(T)}_X$ is aimed at curing problems at the boundaries), its shows a very erratic behavior with acute peaks/troughs and even explosion (at the right boundary). This latter feature is easily explained: the right tail of $\hat{f}_S$ is noticeably thicker than that of the genuine Normal distribution, which obviously get their ratio, i.e.\ $\hat{f}^{(T)}_X$ from (\ref{eqn:ratiodens}), to tend to $\infty$ as $x\to 1$. 

\ppn The phenomenon at the left boundary is also typical of what may be observed with such an estimate. In fact, $\hat{f}_S$ is slightly rougher than $\phi$ and slowly meanders around it. Although quite discreet and not much disturbing on the $S$-scale, these fluctuations are magnified greatly by the back-transformation, especially those in the tails of $f_S$ (when $\phi$ takes on small values in (\ref{eqn:ratiodens})). Moreover, in the $X$-domain, the frequency of the fluctuations gets higher and higher when approaching the boundaries. Of course, this is again the straight effect of the back-transformation: the tails of $\hat{f}_S$ are bluntly shrunk back into $[0,1]$, resulting in the supremely wiggly behavior of $\hat{f}_X^{(T)}$ close to 0 (the first trough is so acute that it is barely visible). The $X$-scaling on top of the left panel, which gives the standard normal quantile for the corresponding $S$-value, allows one to contemplate the amount of that shrinkage. In short, the back-transformation does 
not allow the homogeneous smoothness of the estimate on the $S$-scale to carry over to the $X$-scale. As is, this estimate (\ref{eqn:naivetransf}) is not acceptable, indeed, like \cite{Jones07b} deplored.

\subsection{An improved probit-transformation kernel density estimator based on local likelihood estimation}\label{subsec:improv}

The foregoing discussion, however, inspires some improvements for the probit-transformation estimator. First, focusing on the estimation of $f_S$ without the final estimate of $f_X$ in mind is pointless, as a good estimator for $f_S$ does not automatically becomes a good estimator for $f_X$. More specifically, it is clear that a global bandwidth on the $S$-scale cannot produce an estimate of $f_X$ which is reasonably smooth all over it support. Rather, it seems sensible to work with a local bandwidth on the $S$-scale. Note that this goes against the initial motivation for using transformations in kernel density estimation: \cite{Wand91} primarily suggested using a transformation to be allowed to conveniently use standard kernel estimation with global bandwidth on the transformed scale, when this was not effective on the initial one. Bandwidth issues are discussed in details in Section \ref{sec:bandwidth}.

\ppn On another note, as exemplified above, it is particularly important that the tails of $f_S$ are estimated smoothly and, of course, as accurately as possible. Yet, standard kernel density estimators are known to do poorly in the tails (`spurious bumps'), so it is clear that (\ref{eqn:naivedensS}) is not the best choice for estimating $f_S$ here. In fact, the smoothness of $\hat{f}_S$ should be guided so that it is similar to the smoothness of $\phi$ all over the real line, in particular in the tails. 
Here, ``same smoothness as $\phi$'' has to be understood not in terms of $\hat{f}_S$ being infinitely many times differentiable (this is known to depend on the kernel $K$ only), but an estimate $\hat{f}_S$ not spuriously fluctuating much around $\phi$ anywhere. An easy way to achieve this is to ask $\hat{f}_S$ to locally behave like $\phi$, and this naturally leads to estimating $f_S$ via local likelihood methods \citep{Loader96,Hjort96,Park02}.

\ppn \cite{Hjort96}'s local likelihood approximates locally the unknown $f_S$ by a density from a given parametric family, for which the Gaussian family is the obvious choice here. Now, given that $\phi(s) = (2\pi)^{-1/2}\exp(-s^2)$, this is also equivalent to locally approximating $\log f_S$ by a constrained polynomial of degree 2. Locally fitting a polynomial to the log-density is actually \cite{Loader96}'s formulation of local likelihood estimation, and has some advantages over local parametric density estimation: the practical implementation is easier (closed-form expressions often exist for the estimators) and the asymptotic theory is more transparent (the bias does not depend on a `best local parametric approximant' which is hard to define in practice). Therefore, only \cite{Loader96}'s local log-polynomial density estimation will be used here for $f_S$, although it is stressed that the motivation for that was originally to be found in local Gaussian estimation. In any case, final estimates using 
either formulation of local likelihood should be comparably similar. In particular, both are known to have favorable tail behavior and to be able to correctly make it up for the derivatives of the underlying density. Therefore, local likelihood estimation really meets the needs exposed above. 

\ppn The local log-polynomial likelihood method assumes that, around $s$, the log-density can be well approximated by a polynomial of some order, say $p$: $\log f_S(t) \simeq a_0(s) + a_1(s) (t-s) +\ldots + a_p(s) (t-s)^p$. The local coefficients are then found by solving a weighted maximum likelihood problem
\begin{multline} (\tilde{a}_0(s),\ldots,\tilde{a}_p(s)) = \arg \max_{a_0,a_1,\ldots,a_p}\left\{ \sum_{i=1}^n K\left(\frac{S_i-s}{h}\right) \left(a_0 + a_1 (S_i-s) +\ldots + a_p (S_i-s)^p \right)\right.\\ \left.- n \int_\R K\left(\frac{t-s}{h}\right)\left(a_0 + a_1 (t-s) +\ldots + a_p (t-s)^p\right)\,dt\right\}. \label{eqn:loclikpol} \end{multline}
The estimate of $f_S$ at $s$ is then defined as $\tilde{f}^{(p)}_S(s) = \exp(\tilde{a}_0(s))$, which finally yields, via (\ref{eqn:ratiodens}), $\tilde{f}_X^{(Tp)}(x) = \tilde{f}^{(p)}_S(\Phi^{-1}(x))/\phi(\Phi^{-1}(x))$, an estimate of $f_X$ at any $x \in (0,1)$. Following the idea of local Gaussian estimation, only $p=1$ or $p=2$ will be considered in (\ref{eqn:loclikpol}). This yields two {\it improved probit-transformation kernel density estimators}: $\tilde{f}_X^{(T1)}(x)$ (based on local log-linear estimation of $f_S$, $p=1$) and $\tilde{f}_X^{(T2)}(x)$ (based on local log-quadratic estimation of $f_S$, $p=2$). In the next section, the asymptotic properties of these two estimators are derived, after those of the naive probit-transformation kernel density estimator (\ref{eqn:naivetransf}). 

\section{Asymptotic properties}\label{sec:asymptotics}

The kernel $K$, in (\ref{eqn:naivetransf}) as well as in (\ref{eqn:loclikpol}), will be assumed throughout to be the Gaussian kernel, as this seems to be the obvious choice in this framework. For that kernel, $\int u^2 K(u)\,du = 1$ and $\int K^2(u)\,du = \frac{1}{2\sqrt{\pi}}$. These quantities frequently arise in the properties of kernel density estimators, and direct use of those particular values will be made in the following results.

\subsection{The naive probit-transformation kernel density estimator} \label{subsec:theonaive}

From (\ref{eqn:ratiodens}), it is clear that, for any $x\in (0,1)$, 
\begin{equation} \E\left(\hat{f}^{(T)}_X(x)  \right) = \frac{\E\left(\hat{f}_S(\Phi^{-1}(x)) \right)}{\phi(\Phi^{-1}(x))} \qquad \text{ and } \qquad  \var\left(\hat{f}^{(T)}_X(x)  \right) = \frac{\var\left(\hat{f}_S(\Phi^{-1}(x)) \right)}{\phi^2(\Phi^{-1}(x))}. \label{eqn:EfhatX} \end{equation}
The properties of the conventional estimator $\hat{f}_S$ (\ref{eqn:naivedensS}) are well known. If $f_S$ is twice continuously differentiable at $s \in \R$, then 
\begin{equation} \E\left(\hat{f}_S(s)\right) = f_S(s) + \frac{1}{2}h^2 f''_S(s) +o(h^2) \qquad \text{ and } \qquad \var\left(\hat{f}_S(s)\right) = \frac{f_S(s)}{2nh\sqrt{\pi}} +o((nh)^{-1}) \label{eqn:biasfS} \end{equation}
if $h \to 0$ and $nh \to \infty$ as $n \to \infty$, conditions which consequently guarantee the consistency of $\hat{f}_S$. 

\ppn From $f_S(s) = f_X(\Phi(s))\phi(s)$ and using $\phi'(s) = -s\phi(s)$ and $\phi''(s) = (s^2-1)\phi(s)$, one obtains upon differentiation
\vspace{-0.5cm}
\begin{align}  
f'_S(s) & = f'_X(\Phi(s))\phi^2(s) -s f_X(\Phi(s))\phi(s)  \label{eqn:fSprime} \\ 
\text{ and } \qquad f''_S(s) &= f''_X(\Phi(s)) \phi^3(s) - 3sf'_X(\Phi(s))\phi^2(s) + (s^2-1)f_X(\Phi(s))\phi(s). \label{eqn:fSsecond}
\end{align}
Given that $\phi$ and $\Phi$ are infinitely many times differentiable on $\R$, $f_S$ is seen to be twice continuously differentiable at $s$ if and only if $f_X$ is twice continuously differentiable at $\Phi(s)$. Plugging this in (\ref{eqn:biasfS}) and then in (\ref{eqn:EfhatX}) yields, for any fixed $x \in (0,1)$ at which $f_X$ is twice continuously differentiable: \begin{multline}
\E\left(\hat{f}^{(T)}_X(x)  \right) = f_X(x)\\ +\frac{1}{2}h^2 \left\{ f''_X(x) \phi^2(\Phi^{-1}(x)) - 3f'_X(x)\Phi^{-1}(x)\phi(\Phi^{-1}(x)) + (\{\Phi^{-1}(x)\}^2-1)f_X(x)\right\} +o(h^2), \label{eqn:biasnaive}
\end{multline}
\begin{equation} \text{ and } \quad \var\left(\hat{f}^{(T)}_X(x)  \right) = \frac{f_X(x)}{2nh\phi(\Phi^{-1}(x))\sqrt{\pi}} +o((nh)^{-1}).  \label{eqn:varnaive} \end{equation}
Now, as $\Phi^{-1}(x)$ and $1/\phphx$ grow unboundedly as $x$ approaches 0 or 1, these results have to be modified when considering sequences of values $x_n \to 0/1$. For a sequence such that $x_n/h^m \to \eta$ or $(1-x_n)/h^m \to \eta$ as $n \to \infty$ for some $m,\eta >0$, the asymptotic bias actually becomes
\begin{equation} \text{bias}\left(\hat{f}^{(T)}_X(x_n)  \right) \sim Cmh^2\log h^{-1} \,f_X(x_n) \label{eqn:biasnaivebound} \end{equation}
for some constant $C$, and the asymptotic variance 
\begin{equation} \var\left(\hat{f}^{(T)}_X(x_n)  \right) \sim \frac{f_X(x_n)}{nh^{1+2m}\sqrt{2}\eta^2}, \label{eqn:varnaivebound} \end{equation}
from the limit $(\Phi^{-1}(x_n))^2/(-2\log x_n) \to 1$ as $x_n \to 0$ (and similar for $x_n \to 1$). It turns out that (\ref{eqn:varnaive})/(\ref{eqn:varnaivebound}) is exactly the asymptotic variance of the Gaussian Copula kernel estimator \citep{Jones07a,Jones07b}. Like $\fGC$ and $\fC$, $\fhatTX$ suffers from that unusual boundary effect described in \cite{Jones07a,Jones07b} and \cite{Leblanc12}, namely an increase in the variance order over a small region close to the boundary. The bias of $\fhatTX$, as seen from (\ref{eqn:biasnaive}), is also of similar form to the asymptotic bias of $\fGC$, except for the third term, viz.\ $(\{\Phi^{-1}(x)\}^2-1)f_X(x)$, which is responsible for the boundary bias of higher order $O(h^2\log h^{-1})$ described by (\ref{eqn:biasnaivebound}). Of course, the problem comes from the unbounded nature of $\Phi^{-1}(x)$ as $x \to 0/1$, and it appears that only densities $f_X$ tending to 0 very smoothly as $x$ approaches the boundaries will be correctly estimated there by $\fhatTX$. Otherwise, the estimator is prone to exploding, like what was seen in Figure \ref{fig:naivedens} (right). 

\ppn Expression (\ref{eqn:biasnaive}), however, suggests an easy way to fix this, at least asymptotically. Indeed, instead of (\ref{eqn:ratiodens}), define the amended estimator
\[\fhatTXstar(x)=\frac{\hat{f}_S(\Phi^{-1}(x))}{\phi(\Phi^{-1}(x))\left(1+\frac{1}{2}h^2 (\{\Phx\}^2-1)\right)}. \]
Then, it is clear that $\E\left(\hat{f}^{*(T)}_X(x)\right) = \E\left(\hat{f}^{(T)}_X(x)  \right) /(1+\frac{1}{2}h^2 (\{\Phx\}^2-1))$,
and given that 
\begin{equation} \frac{1}{1+\frac{1}{2}h^2 (\{\Phx\}^2-1)} = 1-\frac{1}{2}h^2 (\{\Phx\}^2-1)+o(h^2)  \label{eqn:biasamend} \end{equation}
as $h \to 0$, it can easily be seen that the amendement exactly makes it up for the the third term in (\ref{eqn:biasnaive}), leading to
\begin{equation} \E\left(\hat{f}^{*(T)}_X(x)\right) = f_X(x)+\frac{1}{2}h^2 \left\{ f''_X(x) \phi^2(\Phi^{-1}(x)) - 3f'_X(x)\Phi^{-1}(x)\phi(\Phi^{-1}(x))\right\} +o(h^2). \label{eqn:biasstar} \end{equation}
The asymptotic variance is unaffected by this amendment, and remains equal to (\ref{eqn:varnaive}) (or its boundary version (\ref{eqn:varnaivebound})). The expectation of $\hat{f}^{*(T)}_X(x)$ is now very similar to that of $\fGC(x)$, which is \citep{Jones07a,Jones07b}
\begin{equation} \E\left(\fGC(x)\right) = f_X(x)+\frac{1}{2}h^2 \left\{ 2f''_X(x) \phi^2(\Phi^{-1}(x)) - 4f'_X(x)\Phi^{-1}(x)\phi(\Phi^{-1}(x))\right\} +o(h^2). \label{eqn:biasGC} \end{equation}
Neither is uniformly smaller (in absolute value) than the other, however (\ref{eqn:biasstar}) seems to get an edge over (\ref{eqn:biasGC}). For instance, at $x=0.5$ (there $\Phi^{-1}(x)=0$), the bias expression of $\fGC$ just shows an extra factor 2. Sensible values for the bandwidth $h$ in both expressions may be quite different, though, so that the comparison is not that straightforward. However, reproducing the analysis shown in \citet[Section 4.4]{Jones07b} allows one to find that the `multiplier' of the optimal bandwidth for $\fhatTXstar$ is $h_0= 2.5679$, whereas it was 2.049 for $\fC$ and 1.946 for $\fGC$. Hence, in terms of asymptotic bias at $x=0.5$, $\fhatTXstar$ does slightly better than $\fGC$ but slightly worse than $\fC$. Further, the `multiplier' for the optimal Mean Squared Error of $\hat{f}^{*(T)}_X$ is found to be $(64\pi^2)^{-1/5}$, that is, {\it exactly} what \cite{Jones07b} found for the multiplier of the MSE of the Beta kernel estimator at $x=0.5$ (and consequently smaller than that of 
$\fGC$). Asymptotically, this amended probit-transformation estimator therefore appears to be a serious competitor for the other two. In practice, though, good performance for it is not guaranteed, in particular close to the boundaries: there, $\{\Phi^{-1}(x)\}^2$ is large, and the linear approximation (\ref{eqn:biasamend}) is dubious. Thus, the amendment may not bring the expected bias correction. Also, the amendment implies that $\hat{f}^{*(T)}_X$ does not integrate to one any more, which calls for a renormalization such as $\hat{f}^{*(T)}_X(x) \leftarrow \hat{f}^{*(T)}_X(x)/\int_0^1 \hat{f}^{*(T)}_X(x) \,dx$. Such a renormalization is, however, commonplace in other nonparametric density estimation procedures and is not an issue {\it per se}.

\ppn Finally, it is noted that the similarities in behavior of the probit-transformation estimator (at least its amended form) and \cite{Jones07a,Jones07b}'s estimator $\fGC$, as evidenced by their almost identical asymptotic bias and variances, is not surprising. As said earlier, normal densities in the $S$-scale transform into conditional densities from bivariate Gaussian copulas in the $X$-scale. Now, if the estimator (\ref{eqn:naivedensS}) is thought of as the sum of Gaussian bumps kernel estimators are usually understood as, then back into the $X$-scale, those bumps become the `Gaussian Copula' bumps the estimator $\fGC$ is constructed from \citep[Figure 3]{Jones07b}.

\subsection{Improved probit-transformation kernel density estimators}

Now the properties of the `improved' probit-transformation estimators of $f_X$ based on local log-polynomial estimation of $f_S$ are derived, starting with $p=1$ in (\ref{eqn:loclikpol}).
\citet[Section 5.2]{Hjort96} derived a closed-form expression for that estimator, and concluded, in accordance with \cite{Loader96}, that, if $f_S$ is twice continuously differentiable at $s$ and $f_S(s)>0$,
\begin{equation} \E\left(\tilde{f}_S^{(1)}(s)\right)  = f_S(s) + \frac{1}{2}h^2\left(f''_S(s) - \frac{f'^2_S(s)}{f_S(s)} \right) +o(h^2)  =  f_S(s) + \frac{1}{2}h^2f_S(s) \left(\log f_S(s)\right)'' +o(h^2) \label{eqn:EfS1} \end{equation}
\begin{equation}  \text{ and } \qquad  \var\left(\tilde{f}_S^{(1)}(s) \right) = \frac{f_S(s)}{2nh\sqrt{\pi}}+o((nh)^{-1}). \label{eqn:varfS1} \end{equation}
Now, defining the corresponding estimator for $f_X$: $\tilde{f}^{(T1)}_{X}(x) = \tilde{f}_S^{(1)}(\Phi^{-1}(x))/\phi(\Phi^{-1}(x))$,
and acting as in Section \ref{subsec:theonaive} using (\ref{eqn:fSprime}) and (\ref{eqn:fSsecond}) in (\ref{eqn:EfS1}) and (\ref{eqn:varfS1}), it follows that, at any fixed $x \in (0,1)$ such that $f_X(x)$ is twice continuously differentiable and $f_X(x)>0$,
\begin{align} \E\left(\tilde{f}^{(T1)}_{X}(x)  \right) & = f_X(x) + \frac{1}{2}h^2 \left\{\left(f''_X(x)-\frac{f'^2_X(x)}{f_X(x)}\right) \phi^2(\Phi^{-1}(x)) - f'_X(x) \Phi^{-1}(x) \phphx -f_X(x)   \right\} + o(h^2) \label{eqn:biasfT1} \\ 
& = f_X(x) + \frac{1}{2}h^2 f_X(x) \left\{\left(\log f_X(x)\right)'' \phi^2(\Phi^{-1}(x)) -  \left( \log f_X(x)\right)' \Phi^{-1}(x)\phphx -1   \right\} + o(h^2) \notag \end{align}
\begin{equation} \text{ and } \qquad \var\left(\tilde{f}^{(T1)}_{X}(x) \right)  = \frac{f_X(x)}{2nh \phphx \sqrt{\pi}}+o((nh)^{-1}). \label{eqn:varfT1} \end{equation}
The variance is the same as that of the naive version of the estimator (and indeed it is again as in (\ref{eqn:varnaivebound}) for sequences of $x$-values converging to one of the boundaries polynomially in $h$), but the bias expression is noticeably different. First, it now mainly depends on the derivatives of the log-density, which is a characteristic of local log-polynomial likelihood estimation. Second and most importantly, it is free from the boundary effect materialized by (\ref{eqn:biasnaivebound}) (it is stressed that the function $\Phi^{-1}(x)\phi(\Phi^{-1}(x))$ appearing in (\ref{eqn:biasfT1}) is bounded). Consequently, (\ref{eqn:biasfT1}) also holds for $x \to 0/1$, provided $f_X(x)>0$. Problems at $x$ such that $f_X(x)=0$ are well known for local log-likelihood methods, since the singularity of the log-density cannot be appropriately approximated by local polynomials. Yet, when such a situation arises at one of the boundaries, the problem is actually somewhat mitigated by the transformation, 
compared to raw local log-polynomial likelihood estimation of $f_X$, as $\Phi^{-1}(x)\phi(\Phi^{-1}(x))$ and $\phi^2(\Phi^{-1}(x)) \to 0$ as $x \to 0/1$. Finally, it is noted that, in general, $\tilde{f}^{(T1)}_{X}$ (and $\tilde{f}^{(T2)}_{X}$ below) will not integrate to one. This problem will, however, vanish asymptotically and in any case can be corrected for in practice using a simple renormalization as the one suggested in the previous subsection.

\ppn The second improved probit-transformation kernel density estimator is obtained when locally fitting a second order polynomial for the log-density, that is, taking $p=2$ in (\ref{eqn:loclikpol}).
Again, \cite{Hjort96} provided a closed-form expression for this estimator. They also pointed out that, as for local polynomial regression \citep{Fan96}, locally fitting a second order polynomial decreases the order of the asymptotic bias from $O(h^2)$ to $O(h^4)$, provided that $f_S$ has four continuously differentiable derivatives at $s$. Then, if $f_S(s)>0$, adapting the results of \cite{Loader96} yields
\begin{align*} \E\left(\tilde{f}^{(2)}_{S}(s)\right) & = f_S(s) - \frac{1}{8} h^4 f_S(s) \left((\log f_S(s))''''+ 4 (\log f_S(s))'''(\log f_S(s))' \right) +o(h^4)  \label{eqn:EfS2}  
\\  & = f_S(s) - \frac{1}{8} h^4 \left(f''''_S(s) -3\,\frac{f''^2_S(s)}{f_S(s)} + 2 \, \frac{f'^4_S(s)}{f^3_S(s)} \right) +o(h^4) \notag \end{align*}
\begin{equation} \text{ and } \qquad \var\left(\tilde{f}^{(2)}_{S}(s) \right) = \frac{27}{16}\,\frac{f_S(s)}{2nh\sqrt{\pi}}+o((nh)^{-1}). \label{eqn:varfS2} \end{equation}
(\ref{eqn:fSsecond}) can be differentiated further to obtain the first four derivatives of $f_S$ in terms of $f_X$ and $\phi$, and one gets for any fixed $x \in (0,1)$ such that $f_X(x)$ is positive and four times continuously differentiable:
\begin{align} \E\left(\tilde{f}^{(T2)}_{X}(x)  \right)  = f_X(x) - \frac{1}{8}h^4 \bigg\{  &\left(f''''_X(x)-3\frac{f''^2_X(x)}{f_X(x)}+2\frac{f'^4_X(x)}{f^3_X(x)}\right)\phi^4(\Phi^{-1}(x)) \notag \\ & +2\left(9\,\frac{f'_X(x)f''_X(x)}{f_X(x)}-4\,\frac{f'^3_X(x)}{f^2_X(x)}-5f'''_X(x) \right)\Phx\phi^3(\Phi^{-1}(x)) \notag  \\ & +\left(\left(19\left\{\Phi^{-1}(x)\right\}^2-4\right)f''_X(x) - 15\left\{\Phi^{-1}(x)\right\}^2 \frac{f'^2_X(x)}{f_X(x)} \right)\phi^2(\Phi^{-1}(x)) \notag  \\ & +\left(7\Phi^{-1}(x) - 5\left\{\Phi^{-1}(x)\right\}^3\right)f'_X(x)\phphx\bigg\} + o(h^4)  \label{eqn:biasfT2} \end{align}
\begin{equation} \text{ and } \qquad \var\left(\tilde{f}^{(T2)}_{X}(x)  \right)  = \frac{27}{16}\,\frac{f_X(x)}{2nh\phphx\sqrt{\pi}}+o((nh)^{-1}). \label{eqn:varfT2}\end{equation}
Compared to that of $\tilde{f}^{(T1)}_{X}(x)$, the variance has been inflated by a factor $27/16$, which is the usual variability inflation factor from local linear to local quadratic regression estimators when a Gaussian kernel is used \citep[Section 3.3.1]{Fan96}. Similarly to (\ref{eqn:varnaivebound}), for a sequence of values $x_n$ tending to 0 or 1 such that $x_n/h^m \to \eta$ or $(1-x_n)/h^m \to \eta$ for some $m,\eta >0$, the asymptotic variance is $\var\left(\tilde{f}^{(T2)}_{X}(x_n)\right) \sim 27/16\,f_X(x_n)/(nh^{1+2m}\sqrt{2}\eta^2)$. On the other hand, the bias is now of order $O(h^4)$, as expected, and its expression involves the first four derivatives of $f_X$. The important point, though, is that, like (\ref{eqn:biasfT1}), this holds also for $x \to 0/1$ (provided $f_X(x)>0$) with no extra boundary effect. Besides, this leading bias term even tends to 0 as $x$ tends to 0 or 1, since the functions $\left\{\Phi^{-1}(x)\right\}^r\phi^s(\Phi^{-1}(x))$ tend to 0, for $r,s$ integers, $r \geq 0$, $s 
> 0$. This was not the case for $\tilde{f}^{(T1)}_{X}(x)$, due to the third term in the bracket in (\ref{eqn:biasfT1}). Therefore, at the boundary, the bias of $\tilde{f}^{(T2)}_{X}(x)$ is actually $o(h^4)$, and this time the boundary effect appears advantageous. This is obviously true if $f_X(x)$ does not tend to 0 too quickly and its derivatives do not grow too quickly to $\infty$ when approaching the boundaries (i.e.\ if no terms in (\ref{eqn:biasfT2}) grows unboundedly there).

\ppn A natural question is what happens when fitting a local log-quadratic estimator if $f_X$ is not four times continuously differentiable. 
Remarkably, one still achieves some bias reduction, compared to $\tilde{f}^{(T1)}_{X}(x)$, under the same smoothness assumption of two continuous derivatives. This can first be understood through \cite{Loader96}'s own words (p.\ 1612): ``(...) {\it the estimate will perform well; existence or otherwise of density derivatives is incidental}''. This fact will carry over to the estimate of $f_X$, and in fact, upon inspection of \cite{Loader96}'s proofs and the previous developments, it can be seen that in this case, provided that $f_X(x)>0$, $\E\left(\tilde{f}^{(T2)}_{X}(x)  \right)  = f_X(x) + o(h^2)$.

\ppn It must be mentioned here that {\it ad hoc} bias reduction (from order $O(h^2)$ to $O(h^4)$) techniques based on multiplicative adjustment, and the resulting reduction of Mean Squared Error rate from the usual $O(n^{-4/5})$ to $O(n^{-8/9})$, were attempted for \cite{Chen99}'s Beta kernel estimators in \cite{Hirukawa10}, but that author observed that his simulation studies (p.\ 480) ``{\it do not necessarily support superiority of the modified beta kernel} [i.e. the bias corrected version] {\it over the beta kernel}''. Contrariwise, combining probit transformation and local log-quadratic estimation achieves that asymptotic bias and MSE reduction in an automatic way (provided that $f_X$ is smooth enough, which was of course also assumed in \cite{Hirukawa10}), and is effective in practice, as evidenced in Section \ref{sec:sim}.

\subsection{On the interest of the transformation}\label{subsec:intrans}

Local likelihood estimation of $f_S$ emerged naturally in Section \ref{subsec:improv} as it mostly addressed the drawbacks of the naive implementation of the probit-transformation idea. This said, transforming the data in the first place mainly aimed at dealing with boundary bias, and local likelihood estimators on their own have been advertised as having advantageous bias behaviors compared to conventional kernel estimators anyway \citep{Loader96,Hjort96}. Hence, the necessity of the transformation in this context can be questioned. There are, however, some reasons for persisting with the transformation idea even if the main estimation step is articulated around local likelihood methods. 

\ppn First, one cannot but notice that local likelihood estimation is far from imposing itself as the panacea for boundary bias correction, as evidenced by the large amount of other treatments still being suggested in the literature. Second, local likelihood estimation has been somewhat questioned in \cite{HallTao02}, who showed among other things that, for densities in $\mathcal{C}_2(\R)$, the MISE of the conventional kernel density estimator is at most that of the analog local log-linear estimator, and can be dramatically lower. Conclusions of this type are not that obvious when comparing integrated squared versions of (\ref{eqn:biasnaive}) or even (\ref{eqn:biasamend}) to (\ref{eqn:biasfT1}). Therefore, it seems heuristically that passing local likelihood methods through a transformation is somewhat beneficial to them. Finally, the transformation idea on its own is known to have numerous other advantages, which motivated its introduction in the first place \citep{Devroye85,Wand91}. The beneficial effect 
of the probit-transformation clearly appears through Section \ref{sec:sim}'s simulation study.


\section{Bandwidth selection}\label{sec:bandwidth}

\subsection{Fixed bandwidth}

As in any nonparametric method, the smoothing parameter, in this case the bandwidth $h$ for estimating $f_S$ via (\ref{eqn:loclikpol}), plays a crucial role in how well the estimator performs. A theoretically optimal value of $h$ can easily be derived by balancing the integrated asymptotic squared bias and variance of the considered estimator ((\ref{eqn:biasfT1})-(\ref{eqn:varfT1}) or (\ref{eqn:biasfT2})-(\ref{eqn:varfT2})). A plug-in approach, aiming at estimating the unknown quantities in that expression, seems however unsuitable here, given the complicated nature of the bias expressions, especially (\ref{eqn:biasfT2}).

\ppn Therefore, an approach based on least-squares cross-validation ideas seems preferable: for the estimators $\tilde{f}^{(Tq)}_{X}$ ($q=1,2$), this is selecting the bandwidth minimizing an estimated version of
\begin{equation} \text{WISE}(h) = \int_{0}^1 \left( \tilde{f}^{(Tq)}_{X}(x) - f_X(x)\right)^2 w(x) \,dx \label{eqn:WMISE} \end{equation}
for some non-negative weight function $w$. In their reference rule, \cite{Jones07a,Jones07b} considered the weight function $w(x)=\phi(\Phi^{-1}(x))$ when approximating WISE. Within the probit transformation setting, however, it is noticeable that this  corresponds to an unweighted least-squares criterion in the $S$-domain. Indeed, the simple change-of-variable $\Phi^{-1}(x)=s$ yields
\begin{align*} \int_0^1  \left( \tilde{f}^{(Tq)}_{X}(x) - f_X(x)\right)^2 \phphx \,dx & = \int_0^1  \left(\frac{ \tilde{f}^{(q)}_{S}(\Phi^{-1}(x)) - f_S(\Phi^{-1}(x))}{\phphx}\right)^2 \,\phphx \,dx \\ &= \int_{-\infty}^{+\infty} \left( \tilde{f}^{(q)}_{S}(s) - f_S(s)\right)^2\,ds. \end{align*}
With this weight function, the selected bandwidth $h$ is thus the usual cross-validated bandwidth aiming at being optimal for estimating $f_S$. However, as stressed in Section \ref{subsec:improv}, a bandwidth good at estimating $f_S$ may not be so when the real goal is to estimate $f_X$. In fact, $w(x)=\phphx$ down-weighs the contribution from the boundary regions to WISE, and this is again in some contradiction with the original motivation of constructing estimators $\tilde{f}^{(Tq)}_{X}$ performing well at the boundaries. A natural idea may then appear to take $w \equiv 1$ in (\ref{eqn:WMISE}), which amounts to choosing the bandwidth as the one value minimizing $\int_{-\infty}^{+\infty} \left( \tilde{f}^{(q)}_{S}(s) - f_S(s)\right)^2 /\phi(s)\,ds$ in the $S$-domain. It was seen empirically that this choice leads to severe oversmoothing, though, as definitely too much emphasis is put on the tails of $f_S$ then, not to mention that this is not even clear whether the previous integral converges. 

\ppn A good compromise between $w(x)= \phphx$ and $w(x) \equiv 1$ seems to be $w(x) = \phphx/f_X(x)$. Intuitively, this down-weighs the contributions of the boundary areas to WISE only if $f_X$ is large there. In a sense, this put the focus on the relative error rather than on the absolute one. 
Bandwidth selection via weighted cross-validation was studied in practice in the simulations (see Section \ref{sec:sim}), and some success was also achieved by using the weight function $w(x) = \phphx/\sqrt{f_X(x)}$. In the $S$-domain, this corresponds to weights $\omega(s)=\phi(s)/f_S(s)$ and $\omega(s)=\sqrt{\phi(s)/f_S(s)}$, respectively, which is again straightforward to see through the change of variable $\Phi^{-1}(x)=s$. Of course, in practice, $f_S(s)$ needs to be estimated in either $\omega(s)$. For that purpose one can just use the conventional estimator $\hat{f}_S(s)$ with bandwidth selected via direct plug-in \citep{Sheather91}. So, essentially, the bandwidth $h$ is selected via classic least-squares cross-validation for estimating $f_S$, except that a weight function is added in the least-squares criterion. Specifically, 
\begin{equation} h_{\text{WLSCV}} = \arg \min_{h >0} \left\{ \int_{-\infty}^{+\infty} \left\{\tilde{f}_S^{(q)}(s)\right\}^2 \hat{\omega}(s) \,ds - \frac{2}{n} \sum_{i=1}^n \tilde{f}_{S(-i)}^{(q)}(S_i)\hat{\omega}(S_i) \right\} \label{eqn:hWLSCV} \end{equation}
where $\hat{\omega}$ denotes the estimated version of the considered weight function $\omega$, and, as usual in cross-validation methods, $\tilde{f}_{S(-i)}^{(q)}$ is the `leave-one-out' version of $\tilde{f}_{S}^{(q)}$ computed on all the observations but $S_i$. Similar ideas of weighted cross-validation have been used in many other contexts before, and cross-validation routines from any software can be used to easily perform this optimization task. 

\ppn It is noted that bandwidth selection via likelihood cross-validation, i.e.\ based on minimizing a Kullback-Leibler divergence rather than a least-squares one, was also initially considered. The estimators $\tilde{f}_S^{(q)}$ being local likelihood estimators, this seems a very natural idea in this setting. The strong links between local likelihood methods and Kullback-Leibler divergence were already underlined in \citet[Section 2]{Hjort96} and \cite{Lee06}. However, the method did not perform competitively against the others, and the results are not shown for sake of brevity.

\subsection{$k$-Nearest-Neighbor bandwidth}

The above discussion focus on a fixed bandwidth $h$, constant over $\R$ when estimating $f_S$. However, as discussed in Section \ref{subsec:improv}, a {\it local} bandwidth seems more appropriate on the $S$-scale for producing homogeneous smoothness for $\tilde{f}^{(Tq)}_X$ over $\Is$ in the $X$-domain. An easy way to achieve local bandwidth estimation, and the adaptive behavior thereof, is to use $k$-Nearest-Neighbor ($k$-NN) methods. For conventional kernel density estimation, though, $k$-NN methods are known to produce extremely rough density estimates with fat tails, and are consequently barely used. That said, it appears that those issues mostly disappear when they are used in conjunction with local likelihood density estimation, as already pointed out in \citet[Section 3.4]{Simonoff96}. Also, \cite{Davison07} suggested a procedure which is essentially close to $k$-NN ideas to remove the bumps in the tails of the kernel density estimator. Hence using a $k$-NN bandwidth seems totally appropriate here. 

\ppn Specifically, the local bandwidth, say $h(s)$, is set to $d_k(s)=|s-S_{(k),s}|$, where $S_{(k),s}$ is the $k$th closest observation to $s$ out of the pseudo-sample $\{S_1,\ldots,S_n\}$. It is now $\alpha = k/n$, the fraction of sample observations that will actively enter the estimation of $f_S(s)$ at any $s$, which will play the role of the smoothing parameter. As $\Phi^{-1}$ is monotonic increasing, this is actually also the fraction of observations from the initial sample $\{X_1,\ldots,X_n\}$ that will actively enter the estimation of $f_X(x)$ at any $x \in \Is$. The smoothing parameter is thus selected as if it was directly in the $X$-domain, hence the appropriateness of a $k$-NN rule here. 
In practice, $\alpha$ can be chosen exactly as in the previous subsection, the minimization in (\ref{eqn:hWLSCV}) now being performed with respect to $\alpha$ instead of to $h$.

\ppn Another insight into why $k$-NN methods can be useful for the probit-transformation estimators is got through the following. For $h$ a fixed bandwidth, the asymptotic variance of $\tilde{f}^{(q)}_S(s)$, $q=1,2$ is $ C_q \times f_S(s)/(2nh \sqrt{\pi})$, with $C_1=1$ and $C_2=27/16$, as given in (\ref{eqn:varfS1}) and (\ref{eqn:varfS2}). On the other hand, $d_k(s)$ is the $k$th order statistics of the `sample' of distances between the observations and $s$, and it has been shown \citep{Mack79} that this is such that $\E\left(1/d_k(s)\right) \simeq  2f_S(s)/\alpha$. Hence it can be seen that, under our assumptions, the variance of $\tilde{f}_S^{(q)}$ with such a $k$-NN bandwidth is asymptotically $\var\left(\tilde{f}^{(K)}_S(s)\right) \simeq C_q f^2_S(s)/(n\alpha\sqrt{\pi})$, provided that $n \alpha \to \infty$ and $\alpha \to 0$. Back in the $X$-domain through (\ref{eqn:EfhatX}), one gets
\begin{equation} \var\left(\tilde{f}^{(Tq)}_X(x)\right) \simeq C_q \times \frac{f^2_X(x)}{n\alpha \sqrt{\pi}}. \label{eqn:varknn} \end{equation}
This variance is now free from any factor $1/\phphx$ which can make (\ref{eqn:varfT1}) and (\ref{eqn:varfT2}) arbitrarily large when $x$ is close to the boundaries. Consequently, (\ref{eqn:varknn}) should hold true as such also for $x \to 0/1$, free from boundary effects formalized by (\ref{eqn:varnaivebound}) or similar. 

\ppn The foregoing argument is only heuristic, though, as the asymptotic variance formulas (\ref{eqn:varfS1}) and (\ref{eqn:varfS2}) are valid only under the usual `nonparametric assumption' that $ h \to 0$ (`small bandwidth asymptotics'). However, local likelihood methods may also act as {\it semiparametric} estimators, when $h \to \infty$ (see the discussion in \cite{Hjort96}), giving rise to another asymptotic behavior (`large bandwidth asymptotics'), see \cite{Eguchi98,Park06} for details. In practice, the limit between small and large bandwidth asymptotics is evidently quite fuzzy, and it is not clear which asymptotic approximation, if any, provides the best picture in a given situation. The question is even more relevant when the bandwidth is essentially random, as in the case of $k$-NN estimation. This is another reason for selecting the smoothing parameter, either $h$ or $\alpha$, via cross-validation, as this is not directly based on asymptotics (unlike the plug-in bandwidth selection rules). In any 
case, the above discussion leading to (\ref{eqn:varknn}) was only meant to further motivate the usage of $k$-NN methods for the probit-transformation estimators. Estimators using a $k$-NN bandwidth will indeed prove superior in most of the situations in the simulations below.

\section{Simulation study}\label{sec:sim}


The practical performance of the procedures exposed in the previous sections is now compared to the Beta kernel estimator $\fC$, the Gaussian Copula kernel estimator $\fGC$ and the diffusion estimator $\fdiff$. For reference, the conventional kernel density estimator (without any correction) and its corrected version given by \cite{Dai10}'s `simple boundary correction' have also been incorporated to the study. No other boundary correction methods have been considered, as \cite{Dai10}'s simulations evidenced that their simple correction was essentially matching the performance of the other, more elaborated methods.

\ppn A thousand independent samples, of size $n=50$ and $n=500$, were simulated from the 16 $[0,1]$-supported densities considered in \cite{Jones07a,Jones07b}, see those papers for their exact description. These densities show a wide range of shapes and features (multimodality, smoothness inhomogeneity, unboundedness at boundaries, cusp) and therefore allows a comprehensive comparison of the estimators under test. Those were used to estimate the different densities from the generated samples on an equally spaced grid of 999 points (from 0.001 to 0.999). For a given estimate, say $\hat{f}$, the estimation accuracy was quantified by the Integrated Squared Error (ISE) approximated on the grid of points, i.e.\ $\widehat{\text{ISE}}(\hat{f}) = \frac{1}{1000} \sum_{i=1}^{999} \left(\hat{f}(i/1000) - f(i/1000)\right)^2$. The Mean Integrated Squared Error (MISE) of the different estimators was finally approximated by averaging the obtained ISE over the 1,000 Monte-Carlo replications. Results can be found in Table \
ref{tab:sim50} for $n=50$ and Table \ref{tab:sim500} for $n=500$. More details on practical issues are given below.


\ppn For the Beta and the Gaussian Copula kernel estimators, $\fC$ and $\fGC$, the bandwidth was selected according to \cite{Jones07a,Jones07b}'s `rule-of-thumb' methods. It must be noted straightaway that this may or may not be suitable. For this type of reference rules, the appropriateness of the selected bandwidth heavily depends on the adequacy of the assumed parametric shape for the true density. Naturally, for $\fC$, a Beta density is used as reference, whereas for $\fGC$ the reference density is  a conditional density from a bivariate Gaussian Copula. In these simulations, the selected bandwidth will therefore be ideal for densities of these types (Densities 1, 3 and 5 for $\fC$, Density 15 for $\fGC$), but may be inappropriate in other situations (typically for the multi-modal Densities 2, 13 and 14). There are apparently no other options in practice, though. \cite{Jones07a} themselves compared the performances of $\fC$ and $\fGC$ at estimating Densities 2, 13 and 14 using that `rule-of-thumb', 
although none would give any reliable estimation of those three densities. \cite{Hirukawa10} does not act differently in his simulation studies, and this is therefore also what is done here. \cite{Botev10}'s diffusion kernel density estimator $\hat{f}_{X,\text{diff}}$ is advertised to come with an automatic bandwidth selection rule, which was obviously used here.

\ppn For the conventional kernel estimator $\hat{f}_X$ and its \cite{Dai10}'s corrected version (called $\hat{f}_{X\text{corr}}$ below), the bandwidth was selected via \cite{Sheather91}'s plug-in method on the raw data set. \cite{Dai10}'s method is just to take as local bandwidth at $x$ the minimum between the global, previously selected bandwidth and the distance from $x$ to the closest boundary, and to apply a renormalization. For the naive probit-transformation kernel estimators, the bandwidth in the $S$-domain was again chosen according to \cite{Sheather91}'s rule on the pseudo-sample $\{S_1,\ldots,S_n\}$. For the improved probit-transformation estimators, all the possibilities described in Sections \ref{sec:bandwidth} were studied, which yielded 12 more estimators: estimators $\tilde{f}^{(T1)}_X$ based on local log-linear estimation of $f_S$ with fixed bandwidth $h$ selected via 1) unweighted Least-Squares Cross-Validation (LSCV), 2) weighted LSCV with weight function $\omega(s)=\sqrt{f_S(s)/\phi(s)}$ (
WLSCV1) and 3) weighted LSCV with weight function $\omega(s)=f_S(s)/\phi(s)$ (WSLCV2); estimators $\tilde{f}^{(T2)}_X$ based on local log-quadratic estimation of $f_S$ with fixed bandwidth $h$ selected via the same 3 methods (LSCV, WLSCV1, WLSCV2); and estimators $\tilde{f}^{(T1)}_X$ and $\tilde{f}^{(T2)}_X$ with $k$-NN bandwidth with the value of $\alpha$ again selected via LSCV, WLSCV1, WLSCV2. In the above weight functions $\omega(s)$, $f_S(s)$ was estimated by (\ref{eqn:naivedensS}) with \cite{Sheather91}'s bandwidth.

\ppn Finally, for assessing the usefulness of the transformation and backing the observations made in Section \ref{subsec:intrans}, the results for the local log-linear and log-quadratic density estimators computed directly on the initial dataset $\{X_1,\ldots,X_n\}$ are also shown. The bandwidth ($h$ or $\alpha$) was selected via least-squares cross-validation (LSCV). Those estimators are denoted $\check{f}_X^{(1)}$ and $\check{f}_X^{(2)}$ below. 

\ppn All the computations were run in the {\tt R} software, using the built-in functions for local log-polynomial density estimation available in the {\tt locfit} package. The practical implementation of the discussed methods is therefore straightforward and available to anybody.

\subsection{Results - $n=50$}

\begin{sidewaystable}[h]
\footnotesize
\centering
\begin{tabular}{||l  || r| r |r | r | r | r | r | r | r | r | r | r | r | r | r | r ||r||}
\hline \hline
 &    \multicolumn{16}{|c||}{Densities} & \\
\hline 
Estimator  & \multicolumn{1}{|c|}1 & \multicolumn{1}{|c|}2 & \multicolumn{1}{|c|}3 & \multicolumn{1}{|c|}4 & \multicolumn{1}{|c|}5 & \multicolumn{1}{|c|}6 & \multicolumn{1}{|c|}7 & \multicolumn{1}{|c|}8 & \multicolumn{1}{|c|}9 & \multicolumn{1}{|c|}{10} & \multicolumn{1}{|c|}{11} & \multicolumn{1}{|c|}{12} & \multicolumn{1}{|c|}{13} & \multicolumn{1}{|c|}{14} & \multicolumn{1}{|c|}{15} & \multicolumn{1}{|c||}{16} & \multicolumn{1}{|c||}{Total} \\
\hline \hline
$\hat{f}_X$           &0.062 &\uline{0.096} &0.134 &0.092 &0.565 &{\bf \uline{0.053}} &0.102 &\uline{0.035} &0.091 &0.550 &0.070  &{\bf \uline{0.093}} &0.312 &0.314 &\uline{0.048}  &{\bf \uline{0.148}} & 0.173 \\
 $\hat{f}_{X\text{,corr}}$ &0.061 &{\bf \uline{0.089}} &0.137 &0.106 &0.284 &\uline{0.065} &0.107 &0.061 &0.110 &0.228 &0.090  &\uline{0.112} &\uline{0.257} &0.291 &\uline{0.055}  &{\bf \uline{0.149}} & 0.138\\
 \hline
$\fC$                   &\uline{0.045} &0.470 &\uline{0.055} &{\bf \uline{0.034}} &0.364 &{\bf \uline{0.053}} &0.082 &\uline{0.034} &\uline{0.039} &0.367 &0.074  &0.216 &0.582 &0.632 &\uline{0.051}  &\uline{0.182} & 0.205 \\
$\fGC$        &\uline{0.054} &0.457 &\uline{0.073} &\uline{0.037} &0.168 &\uline{0.066} &\uline{0.050} &{\bf \uline{0.033}} &\uline{0.064} &\uline{0.144} &\uline{0.043}  &0.296 &0.729 &0.586 &\uline{0.039}  &0.223 & 0.191\\
$\hat{f}_{X\text{,diff}}$                &0.059 &0.622 &{\bf \uline{0.051}} &\uline{0.037} &0.439 &\uline{0.069} &{\bf \uline{0.045}} &\uline{0.037} &{\bf \uline{0.033}} &0.442 &{\bf \uline{0.030}}  &\uline{0.108} &\uline{0.302} &0.633 &0.066  &\uline{0.167} & 0.196 \\
\hline
$\hat{f}_X^{(T)}$                  &0.073 &\uline{0.117} &0.151 &0.126 &0.176 &0.093 &0.118 &0.084 &0.130 &0.167 &0.107  &0.163 &\uline{0.250} &0.372 &0.073  &\uline{0.166} & 0.148\\
$\hat{f}^{*(T)}_X$         &0.071 &\uline{0.114} &0.101 &\uline{0.050} &0.246 &0.083 &\uline{0.068} &\uline{0.055} &0.093 &0.184 &\uline{0.067}  &\uline{0.153} &{\bf \uline{0.215}} &0.365 &0.058  &\uline{0.165} & 0.131 \\
\hline
$\tilde{f}_X^{(T1)}$ - $h$-LSCV         &0.075 &0.179 &0.130 &0.083 &0.148 &0.089 &0.098 &0.075 &0.103 &0.146 &0.091 &21.888 &\uline{0.284} &\uline{0.277} &0.069  &4.903 & 1.790\\
$\tilde{f}_X^{(T1)}$ - $h$-WLSCV1       &\uline{0.053} &0.176 &0.108 &0.075 &\uline{0.120} &0.076 &0.073 &0.072 &\uline{0.078} &{\bf \uline{0.116}} &0.074 &18.214 &\uline{0.250} &\uline{0.278} &0.060  &2.793 & 1.414 \\
$\tilde{f}_X^{(T1)}$ - $h$-WLSCV2       &\uline{0.036} &0.204 &\uline{0.087} &\uline{0.067} &{\bf \uline{0.109}} &0.101 &\uline{0.058} &0.058 &0.290 &{\bf \uline{0.112}} &\uline{0.061} &19.911 &0.738 &0.311 &\uline{0.043}  &2.344 & 1.533\\
$\tilde{f}_X^{(T1)}$ - $\alpha$-LSCV         &0.111 &0.262 &0.145 &0.094 &0.146 &0.106 &0.114 &0.084 &0.120 &0.158 &0.111  &0.186 &0.468 &0.373 &0.090  &0.273 & 0.178\\
$\tilde{f}_X^{(T1)}$ - $\alpha$-WLSCV1        &0.061 &0.242 &0.115 &0.088 &0.126 &\uline{0.067} &0.077 &0.071 &\uline{0.077} &{\bf \uline{0.119}} &0.079  &\uline{0.130} &0.357 &0.356 &0.074  &\uline{0.159} & 0.137 \\
$\tilde{f}_X^{(T1)}$ - $\alpha$-WLSCV2        &\uline{0.039} &0.303 &\uline{0.091} &0.080 &{\bf \uline{0.114}} &{\bf \uline{0.053}} &\uline{0.057} &\uline{0.055} &\uline{0.079} &{\bf \uline{0.118}} &\uline{0.054}  &\uline{0.145} &1.206 &0.370 &0.057  &\uline{0.173} & 0.187\\
\hline
$\tilde{f}_X^{(T2)}$ - $h$-LSCV     &0.372 &0.487 &2.404 &0.655 &0.170 &0.190 &1.285 &1.605 &0.229 &1.092 &1.356 &54.784 &1.748 &0.359 &1.035 &20.034 & 5.488 \\
$\tilde{f}_X^{(T2)}$ - $h$-WLSCV1      &0.186 &0.322 &0.276 &0.223 &{\bf \uline{0.112}} &0.117 &0.830 &0.819 &0.152 &0.348 &0.966 &56.290 &1.259 &0.298 &0.810 &19.004 & 5.126\\
$\tilde{f}_X^{(T2)}$ - $h$-WLSCV2      &0.104 &0.244 &\uline{0.074} &0.091 &{\bf \uline{0.115}} &0.135 &0.656 &0.718 &0.195 &0.191 &0.808 &51.887 &0.771 &\uline{0.270} &0.651 &17.214  & 4.633 \\
$\tilde{f}_X^{(T2)}$ - $\alpha$-LSCV       &\uline{0.053} &0.155 &\uline{0.090} &\uline{0.071} &{\bf \uline{0.111}} &0.080 &\uline{0.066} &\uline{0.048} &0.084 &{\bf \uline{0.118}} &\uline{0.061}  &\uline{0.136} &0.326 &\uline{0.243} &\uline{0.046}  &0.202 & 0.118\\
$\tilde{f}_X^{(T2)}$ - $\alpha$-WLSCV1    &{\bf \uline{0.033}} &\uline{0.141} &\uline{0.073} &\uline{0.071} &{\bf \uline{0.108}} &0.073 &\uline{0.061} &\uline{0.041} &\uline{0.072} &{\bf \uline{0.116}} &\uline{0.051}  &\uline{0.152} &\uline{0.264} &\uline{0.238} &\uline{0.037}  &0.204 & 0.108\\
$\tilde{f}_X^{(T2)}$ - $\alpha$-WLSCV2     &{\bf \uline{0.032}} &\uline{0.137} &\uline{0.071} &\uline{0.069} &{\bf \uline{0.113}} &0.074 &\uline{0.054} &\uline{0.038} &\uline{0.072} &{\bf \uline{0.114}} &\uline{0.046}  &0.348 &\uline{0.252} &\uline{0.237} &{\bf \uline{0.032}}  &0.291 & 0.124 \\
\hline
$\check{f}_X^{(1)}$ - $h$-LSCV         &0.085 &0.166 &0.268 &0.264 &0.670 &0.096 &0.228 &0.084 &0.215 &0.667 &0.177  &0.168 &2.247 &\uline{0.229} &0.100  &0.263 & 0.371\\
$\check{f}_X^{(1)}$ - $\alpha$-LSCV          &0.125 &0.256 &0.371 &0.322 &0.945 &0.149 &0.295 &0.097 &0.316 &0.880 &0.255  &0.255 &0.811 &0.354 &0.123  &0.348 & 0.369\\
$\check{f}_X^{(2)}$ - $h$-LSCV        &0.109 &\uline{0.128} &0.216 &0.154 &0.570 &0.135 &0.203 &0.072 &0.274 &0.578 &0.175  &0.163 &2.044 &{\bf \uline{0.192}} &0.088  &0.393 & 0.343 \\
$\check{f}_X^{(2)}$ - $\alpha$-LSCV         &0.060 &\uline{0.125} &0.248 &0.196 &0.739 &\uline{0.066} &0.203 &0.078 &0.210 &0.723 &0.174  &0.176 &0.396 &\uline{0.210} &0.112  &0.210& 0.245 \\
\hline 
\hline
\end{tabular}
\caption{(estimated) Mean Integrated Squared Error for the 23 estimators under study, when estimating the 16 densities considered in \cite{Jones07a,Jones07b} from samples of size $n=50$. Bold values evidence the minimum MISE for each density (and other MISE's within two standard errors of the minimum MISE), and underlined values show the top 8 for each density (again taking into account possible `ties', i.e.\ non-significantly different values). The column `Total' gives the average MISE over the 16 densities.}
\label{tab:sim50}
\end{sidewaystable}

As seen from Table \ref{tab:sim50}, there is no uniformly best estimator dramatically outperforming all the others for all densities. In fact, 16 different estimators (out of the 23 which are considered) can claim to be best at estimating at least one of the 16 densities. The diffusion kernel estimator $\fdiff$ and the improved probit-transformation estimator $\tilde{f}_X^{(T2)}$ with $\alpha$-WLSCV2 bandwidth lead this particular ranking with 4 first positions each, followed by the conventional estimator $\hat{f}_X$ and $\tilde{f}_X^{(T1)}$ with $\alpha$-WLSCV2 bandwidth, 3 first positions each. On the Mean Integrated Square Error averaged over the 16 densities (column `Total' of Table \ref{tab:sim50}), the three improved probit-transformation estimators $\tilde{f}_X^{(T2)}$ and bandwidth of type $k$-NN lead the way. The exact weight function used in (\ref{eqn:hWLSCV}) is not much important, with a slight preference for the WSLCV1 weight, though. Fourth comes the amended naive probit-transformation 
estimator $\hat{f}^{*(T)}_X$, and fifth $\tilde{f}_X^{(T1)}$ with $\alpha$-WLSCV1 bandwidth.

\ppn If a score of 1 point is given to the estimator ranked first (on MISE) for a particular density, a score of 2 points is given to the estimator ranked second, etc., a `robust' overall ranking may be obtained. Again, the three improved probit-transformation estimators $\tilde{f}_X^{(T2)}$ with $k$-NN bandwidth come first ($\alpha$-WLSCV1, 84 pts; $\alpha$-WLSCV2, 91 pts; $\alpha$-LSCV, 115 pts). The Gaussian Copula kernel estimator is fourth (136 pts) and $\tilde{f}_X^{(T1)}$ with $\alpha$-WLSCV2 bandwidth fifth (139 pts). In this particular ranking, $\fdiff$ is 7th (147 pts) and $\fC$ only 10th (156 pts). 

\ppn Of course, an obvious reason why $\fC$ and $\fGC$ do not perform better on the average is their terrible MISE for the multimodal densities 2, 13 and 14, as a result of using a totally inappropriate bandwidth. As stated in the previous section, however, there is currently no other reliable bandwidth selection rule available in the literature for those two estimators. For instance,  \cite{Chen99} himself sometimes opted for subjective bandwidth selection in his simulations since the devised cross-validation-based procedure was not trustworthy. Bandwidth problems for the Beta kernel estimator are also discussed in \cite{Zhang10}. Besides, the bandwidth is only part of the picture. For Density 1, which is the Beta$(4,4)$ density, the `rule-of-thumb' bandwidth for $\fC$ should be optimal, but $\fC$ is still beaten by the best probit-transformation estimators $\tilde{f}_X^{(T2)}$ (bandwidth $\alpha$-WLSCV1 and $\alpha$-WLSCV2). The same observation holds for $\fGC$ and its optimal bandwidth for Density 15: it 
is also beaten there by $\tilde{f}_X^{(T2)}$ (bandwidth $\alpha$-WLSCV1 and $\alpha$-WLSCV2). It is also noticeable that $\fC$ really struggles at estimating the unbounded Densities 5 and 10 (this is also the case for $\fdiff$). The Gaussian Copula estimator $\fGC$ can reasonably cope with that feature, but is again unable to compete with the improved probit-transformation estimators on it.

\ppn Some other, minor but still noteworthy points are the following. In these small samples, the simple boundary bias correction suggested in \cite{Dai10} is not effective. It manages to reduce the MISE of the conventional estimator for the unbounded densities 5 and 10, for obvious reasons, but for the other cases both estimators are roughly level (the conventional estimator even appears slightly better). On the other hand, the amendment made to the naive probit-transformation estimator seems to effectively improve its MISE. Only for the unbounded densities is the MISE of the initial version smaller than that of the amended version, which is easily understood: the amendment tends to prevent the estimate from growing unboundedly at the boundaries.

\ppn Comparing the improved probit-transformation estimators between them, it is clear that using a bandwidth of type $k$-NN is a must, with this small sample size ($n=50$). The fixed bandwidth versions do not work well, especially for the estimators $\tilde{f}_X^{(T2)}$. For the estimators $\tilde{f}_X^{(T1)}$, the situation is less clear-cut, and for some density shapes a fixed bandwidth may sometimes improve on a $k$-NN bandwidth. For other shapes, however, it may yield very poor results (densities 12 and 16). It is also noticeable that, when a $k$-NN bandwidth is used, local log-quadratic estimation of $f_S$ is preferable over local log-linear estimation, but it is the contrary when a fixed bandwidth is used. Finally, it is clear that the local-likelihood estimators directly targeting $f_X$ without going through the transformation (estimators $\check{f}_X$) cannot compete, as anticipated in Section \ref{subsec:intrans}.

\subsection{Results - $n=500$}

\begin{sidewaystable}[h]
\footnotesize
\centering
\begin{tabular}{||l  || r| r |r | r | r | r | r | r | r | r | r | r | r | r | r | r ||r||}
\hline \hline
 &    \multicolumn{16}{|c||}{Densities} & \\
\hline 
Estimator  & \multicolumn{1}{|c|}1 & \multicolumn{1}{|c|}2 & \multicolumn{1}{|c|}3 & \multicolumn{1}{|c|}4 & \multicolumn{1}{|c|}5 & \multicolumn{1}{|c|}6 & \multicolumn{1}{|c|}7 & \multicolumn{1}{|c|}8 & \multicolumn{1}{|c|}9 & \multicolumn{1}{|c|}{10} & \multicolumn{1}{|c|}{11} & \multicolumn{1}{|c|}{12} & \multicolumn{1}{|c|}{13} & \multicolumn{1}{|c|}{14} & \multicolumn{1}{|c|}{15} & \multicolumn{1}{|c||}{16} & \multicolumn{1}{|c||}{Total} \\ 
\hline \hline
$\hat{f}_X$           &0.010 &{\bf \uline{0.017}} &0.048 &0.036 &0.395 &{\bf \uline{0.009}} &0.039 &\uline{0.009} &0.032 &0.384 &0.025 &{\bf \uline{0.017}} &0.064 &\uline{0.028} &0.010 &\uline{0.029}      &0.072 \\
 $\hat{f}_{X\text{,corr}}$ &0.010 &{\bf \uline{0.016}} &0.026 &0.020 &0.036 &\uline{0.010} &0.021 &\uline{0.009} &0.020 &0.031 &0.016 &\uline{0.017} &\uline{0.052} &\uline{0.027} &0.010 &\uline{0.029}      &0.022\\
 \hline
$\fC$                   &\uline{0.008} &0.155 &{\bf \uline{0.007}} &\uline{0.010} &0.210 &{\bf \uline{0.009}} &\uline{0.011} &{\bf \uline{0.007}} &{\bf \uline{0.005}} &0.217 &0.013 &0.077 &0.226 &0.510 &0.012 &0.061      &0.096\\
$\fGC$        &0.010 &0.226 &\uline{0.013} &\uline{0.013} &0.031 &\uline{0.013} &\uline{0.010} &\uline{0.008} &\uline{0.012} &0.026 &\uline{0.009} &0.135 &0.251 &0.497 &\uline{0.007} &0.076      &0.084\\
$\hat{f}_{X\text{,diff}}$                &0.010 &0.350 &\uline{0.009} &{\bf \uline{0.008}} &0.145 &\uline{0.011} &{\bf \uline{0.008}} &\uline{0.007} &{\bf \uline{0.005}} &0.148 &{\bf \uline{0.005}} &{\bf \uline{0.017}} &0.062 &0.601 &0.016 &\uline{0.030}      &0.090 \\
\hline
$\hat{f}_X^{(T)}$                  &0.011 &\uline{0.020} &0.021 &0.020 &0.027 &0.015 &0.019 &0.012 &0.019 &0.024 &0.016 &0.027 &\uline{0.037} &\uline{0.035} &0.011 &\uline{0.032}      &0.022\\
$\hat{f}^{*(T)}_X$        &0.011 &\uline{0.019} &0.016 &\uline{0.010} &0.076 &0.014 &0.013 &0.010 &\uline{0.015} &0.052 &0.012 &0.027 &{\bf \uline{0.035}} &\uline{0.035} &\uline{0.010} &\uline{0.032}      &0.024 \\
\hline
$\tilde{f}_X^{(T1)}$ - $h$-LSCV         &0.011 &0.025 &0.021 &0.016 &0.025 &0.017 &0.017 &0.013 &0.019 &0.024 &0.015 &0.036 &\uline{0.039} &0.038 &0.011 &0.053      &0.024\\
$\tilde{f}_X^{(T1)}$ - $h$-WLSCV1       &0.011 &0.024 &\uline{0.013} &\uline{0.012} &{\bf \uline{0.013}} &0.014 &0.017 &0.015 &\uline{0.013} &{\bf \uline{0.014}} &0.016 &0.096 &\uline{0.043} &0.037 &0.011 &0.074      &0.027 \\
$\tilde{f}_X^{(T1)}$ - $h$-WLSCV2       &\uline{0.009} &\uline{0.022} &\uline{0.011} &\uline{0.012} &{\bf \uline{0.013}} &0.022 &\uline{0.009} &0.014 &0.020 &\uline{0.017} &\uline{0.010} &0.128 &0.099 &\uline{0.036} &\uline{0.007} &0.114      &0.034\\
$\tilde{f}_X^{(T1)}$ - $\alpha$-LSCV         &0.016 &0.034 &0.022 &0.015 &0.023 &0.017 &0.016 &0.013 &0.018 &0.022 &0.015 &0.025 &0.059 &0.044 &0.013 &0.034      &0.024\\
$\tilde{f}_X^{(T1)}$ - $\alpha$-WLSCV1         &0.012 &0.033 &0.015 &0.015 &\uline{0.014} &\uline{0.012} &0.013 &0.013 &\uline{0.012} &{\bf \uline{0.013}} &0.013 &0.027 &\uline{0.055} &0.044 &0.013 &\uline{0.029}      &0.021\\
$\tilde{f}_X^{(T1)}$ - $\alpha$-WLSCV2        &\uline{0.006} &0.027 &\uline{0.010} &\uline{0.013} &{\bf \uline{0.013}} &\uline{0.013} &{\bf \uline{0.008}} &0.011 &\uline{0.011} &{\bf \uline{0.013}} &\uline{0.008} &\uline{0.018} &1.408 &0.041 &\uline{0.009} &0.057      &0.104\\
\hline
$\tilde{f}_X^{(T2)}$ - $h$-LSCV     &0.060 &0.152 &0.021 &0.016 &0.020 &0.025 &0.316 &0.271 &0.037 &0.021 &0.452 &1.552 &0.642 &0.216 &0.376 &0.350      &0.283 \\
$\tilde{f}_X^{(T2)}$ - $h$-WLSCV1      &0.062 &0.152 &\uline{0.014} &\uline{0.013} &0.018 &0.030 &0.369 &0.299 &0.033 &0.018 &0.485 &1.530 &0.705 &0.223 &0.415 &0.677      &0.315\\
$\tilde{f}_X^{(T2)}$ - $h$-WLSCV2      &0.015 &0.049 &\uline{0.013} &\uline{0.012} &0.018 &0.031 &0.214 &0.262 &0.027 &\uline{0.018} &0.370 &1.266 &0.425 &0.094 &0.295 &0.546      &0.229\\
$\tilde{f}_X^{(T2)}$ - $\alpha$-LSCV       &\uline{0.004} &0.027 &\uline{0.013} &0.015 &{\bf \uline{0.013}} &\uline{0.013} &\uline{0.011} &\uline{0.009} &\uline{0.014} &\uline{0.016} &\uline{0.009} &\uline{0.021} &\uline{0.048} &0.040 &\uline{0.004} &0.040      &0.018\\
$\tilde{f}_X^{(T2)}$ - $\alpha$-WLSCV1    &\uline{0.004} &0.027 &0.014 &0.015 &\uline{0.013} &0.016 &\uline{0.011} &\uline{0.009} &0.021 &\uline{0.016} &\uline{0.009} &\uline{0.020} &0.083 &0.041 &\uline{0.003} &{\bf \uline{0.023}}      &0.020\\
$\tilde{f}_X^{(T2)}$ - $\alpha$-WLSCV2     &{\bf \uline{0.003}} &0.035 &0.015 &0.018 &\uline{0.013} &0.039 &0.012 &\uline{0.009} &0.029 &\uline{0.017} &\uline{0.008} &0.086 &0.089 &0.048 &{\bf \uline{0.003}} &0.292      &0.045 \\
\hline
$\check{f}_X^{(1)}$ - $h$-LSCV          &0.014 &0.026 &0.085 &0.053 &0.243 &0.023 &0.060 &0.041 &0.057 &0.243 &0.047 &0.031 &3.747 &\uline{0.037} &0.055 &0.042      &0.300\\
$\check{f}_X^{(1)}$ - $\alpha$-LSCV          &0.020 &0.037 &0.072 &0.117 &0.229 &0.030 &0.057 &0.045 &0.061 &0.161 &0.048 &0.043 &0.086 &0.045 &0.074 &0.041      &0.073\\
$\check{f}_X^{(2)}$ - $h$-LSCV        &\uline{0.009} &{\bf \uline{0.016}} &0.066 &0.046 &0.245 &0.016 &0.051 &0.026 &0.057 &0.244 &0.039 &\uline{0.020} &0.056 &{\bf \uline{0.025}} &0.025 &0.037      &0.061 \\
$\check{f}_X^{(2)}$ - $\alpha$-LSCV           &0.012 &\uline{0.020} &0.093 &0.062 &0.229 &0.020 &0.066 &0.028 &0.055 &0.162 &0.050 &0.037 &0.075 &\uline{0.033} &0.028 &0.040      &0.063 \\
\hline 
\hline
\end{tabular}
\caption{(estimated) Mean Integrated Squared Error for the 23 estimators under study, when estimating the 16 densities considered in \cite{Jones07a,Jones07b} from samples of size $n=500$. Bold values evidence the minimum MISE for each density (and other MISE's within two standard errors of the minimum MISE), and underlined values show the top 7 for each density (again taking into account possible `ties', i.e.\ non-significantly different values). The column `Total' gives the average MISE over the 16 densities.}
\label{tab:sim500}
\end{sidewaystable}

For larger samples ($n=500$) (see Table \ref{tab:sim500}), the same conclusions essentially hold true, although there are some noteworthy changes. The diffusion estimator $\fdiff$ is now the sole leader in the `first positions' ranking, with 5 first positions. It is followed by the Beta kernel estimator $\fC$ (4 first positions), the conventional estimator $\hat{f}_X$ and $\tilde{f}_X^{(T1)}$ with $\alpha$-WLSCV2 bandwidth (3 first positions). On the average MISE, this is now $\tilde{f}_X^{(T2)}$ with $\alpha$-LSCV bandwidth which is the best, just before the same $\tilde{f}_X^{(T2)}$ with $\alpha$-WLSCV1 bandwidth and $\tilde{f}_X^{(T1)}$ with $\alpha$-WLSCV1 bandwidth. Very closely follow the naive probit-transformation estimator $\hat{f}^{(T)}_X$ and the bias-corrected conventional estimator $\hat{f}_{X,\text{corr}}$. In the ranking induced by the ranks, $\tilde{f}_X^{(T2)}$ - $\alpha$-LSCV is again first (106 pts), ahead of $\tilde{f}_X^{(T1)}$ - $\alpha$-WLSCV2 (114 pts), $\tilde{f}_X^{(T2)}$ - $\alpha$-
WLSCV1 (122 points) and the amended naive probit-transformation estimator $\hat{f}^{*(T)}_X$ (136 points). The diffusion estimator $\fdiff$ (137 pts) completes the Top 5.

\ppn The main point is that $\tilde{f}_X^{(T2)}$ - $\alpha$-WLSCV2 has now disappeared from the top positions. In fact, the WLSCV2 criterion puts a special emphasis on the tails of $f_S$ when estimating it, and this is necessary to achieve smooth estimates of those tails in small samples, when very few observations fall in the tail regions. Contrariwise, in large samples, enough observations are found even in the tails of the density and the `moderate' tail emphasis driven by the criterion WLSCV1, or even no particular emphasis at all (LSCV), works well enough. The other observations made in the previous subsection for $n=50$ remain mostly true for sample of size $n=500$.

\ppn  All in all, without distinction on sample size, the improved probit-transformation estimator $\tilde{f}_X^{(T2)}$ with $k$-NN bandwidth selected by weighted cross-validation (WLSCV1) appears to be the best, overall. It seems therefore reasonable to recommend it as preferred all-around estimator for nonparametrically estimating a density on the unit interval. In small samples, a particular emphasis on the tails of $f_S$, through the weight function WLSCV2, may be appropriate. In fact, the most serious competitor of these estimators seems to be the diffusion kernel estimator $\fdiff$. With its automatic bandwidth selector, it is often doing well, except for unbounded and multimodal densities. The estimators $\tilde{f}_X^{(T2)}$ with $k$-NN bandwidth, however, never do very badly, and often do very well. 

\ppn Actually, they lead many of the rankings one could think of to assess the relative performance of the 23 estimators that were considered. Indeed, the foregoing discussion only considered the Mean Integrated Squared Error, but the same conclusions were drawn by looking only at the boundary behavior of the estimators. For instance, for the same simulations the Mean Squared Error of each estimator at $x=0.01$ and $x=0.99$ was calculated (this analysis again mimics that in \cite{Jones07b}), and the improved probit-transformation estimators were again found to be best on the average, to a similar extent. These results are not shown for sake of brevity, but are available upon request. It is also stressed that the proposed cross-validation bandwidth selection rules showed very consistent behaviors over the Monte-Carlo replications. In other situations, it is well-known that similar LSCV ideas may produce highly variable bandwidths.

\section{Real data example}\label{sec:realdata}

In this section the improved probit-transformation method is used to estimate a probability density from a real data set. The data give the proportion $X$ of white student enrollment in $n=56$ school districts in Nassau County (Long Island, New York), for the 1992-1993 school year. Estimating the density of $X$ has been of interest to assess the common perception in the US in the 90's that public schools were still strongly segregated by race, despite political effort to integrate them. Of course, $X$ being a proportion, its density is supported on $\Is=[0,1]$ and estimating it is somewhat problematic for the reasons explained in Section \ref{sec:intro}. This data set was, among others, considered in \citet[Sections 3.2 and following]{Simonoff96} for illustrating boundary bias problems. 

\begin{figure}[h]
\centering
\includegraphics[scale=0.35]{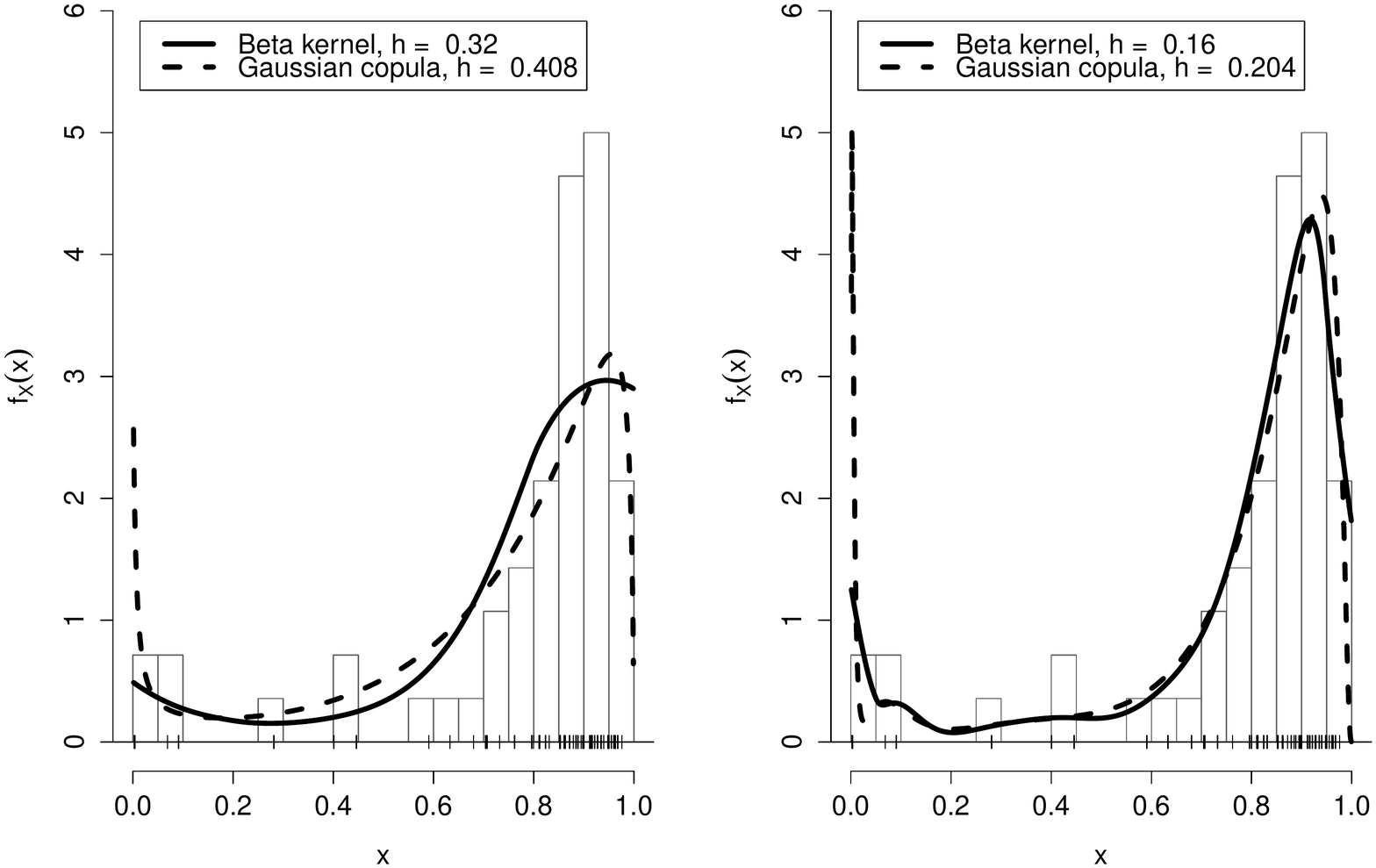}
\caption{Beta kernel (plain) and Gaussian Copula (dashed) estimates with \cite{Jones07a,Jones07b}'s `rule-of-thumb' bandwidths (left panel) and half those (right panel), for the white students enrollment data. These estimates are overlaid to a sample histogram, and the actual observations are displayed as a `rug' along the the bottom of the graph.}
\label{fig:realdat11}
\end{figure}

\ppn Figure \ref{fig:realdat11} (left panel) shows the Beta kernel estimator $\fC$ (with rule-of-thumb bandwidth $h=0.320$) and the Gaussian Copula kernel estimator $\fGC$ (with rule-of-thumb bandwidth $h=0.408$) superimposed on a sample histogram. The Gaussian Copula kernel estimator shows a peak at the 0 boundary, which is actually meaningful here. In fact, two schools showed a white students enrollment extremely close to 0, and the estimator attempts to put a positive probability mass atom at 0, hence the spike. The Beta kernel estimator totally misses this. On the other side, simple visual inspection (see the bottom of the histogram) reveals that there are no observations very close to 1, so the density should reach a maximum at around 0.9-0.95 and then tumble down to 0 when approaching 1. Again, this is mostly what $\fGC$ shows, but the Beta kernel estimator fails to catch this as it remains roughly constant for $x$-values between 0.85 and 1. 

\ppn Note that both estimates appear noticeably oversmoothed, which is not surprising given the clear bimodal nature of the data and the way their bandwidth was selected. Figure \ref{fig:realdat11} (right panel) shows the same estimators but with initial bandwidths divided by two. Such small bandwidths (and even smaller, actually) are necessary to visually recover the curvature of the density for $x$ between 0.5 and 0.9, as suggested by the histogram. The above mentioned two features of the density (spike at 0, drop at 1) are now even more pronounced on the Gaussian Copula kernel estimator whereas the Beta kernel estimator still struggles to evidence them. Both appear quite undersmoothed this time, which makes them visually unpleasant with spurious bumps arising for $x$-values between 0 and 0.4.

\begin{figure}[h]
\centering
\includegraphics[scale=0.35]{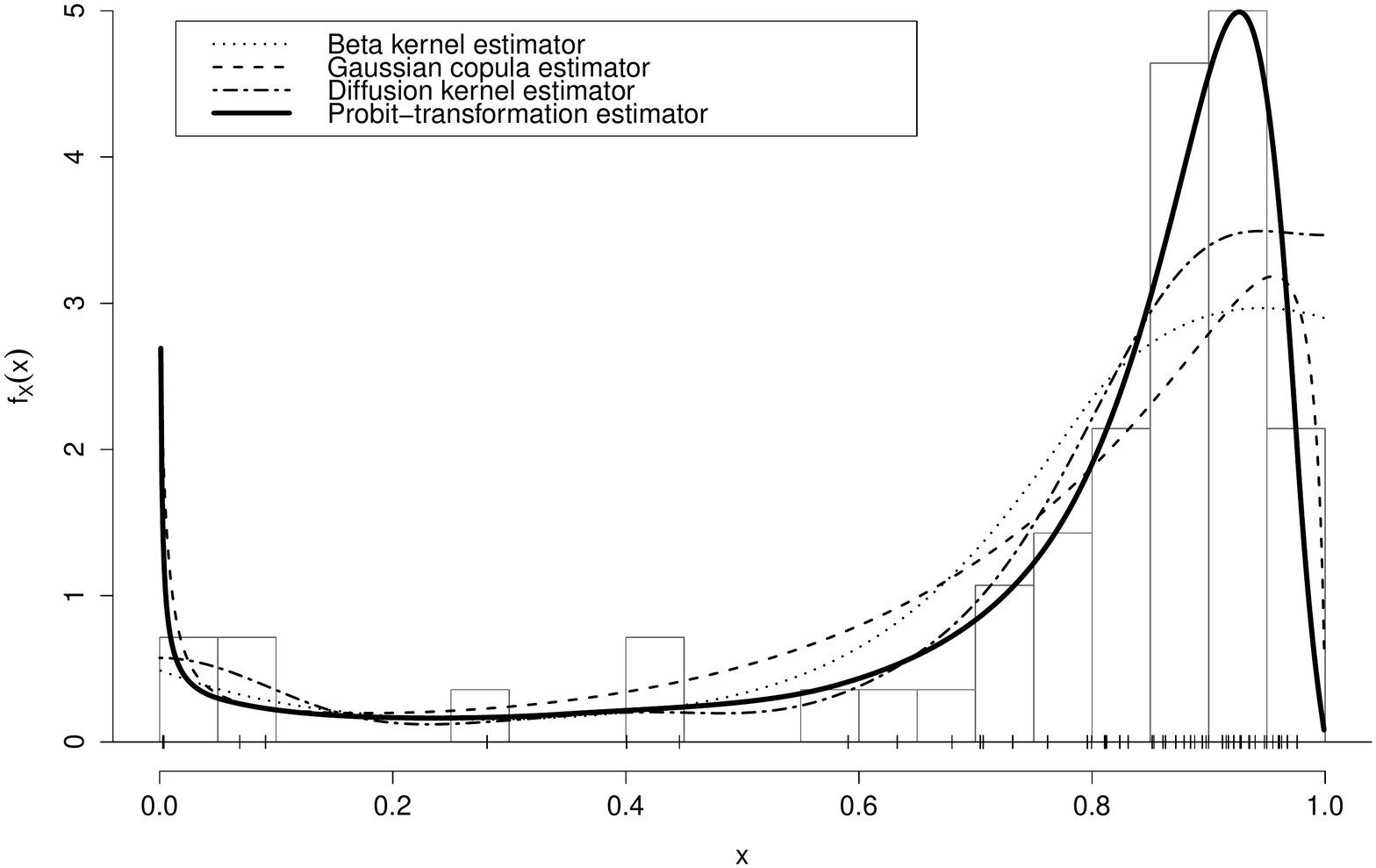}
\caption{Probit-transformation kernel density estimator for the white students enrollment data. Local log-quadratic density estimation was used with a $k$-NN bandwidth selected by weighted least-squares cross-validation ($\alpha=0.885$). Previous Beta and Gaussian Copula kernel estimators with `rule-of-thumb' bandwidths, and the `diffusion' estimator are also shown for easy comparison. }
\label{fig:realdat12}
\end{figure}

\begin{figure}[h]
\centering
\includegraphics[scale=0.3]{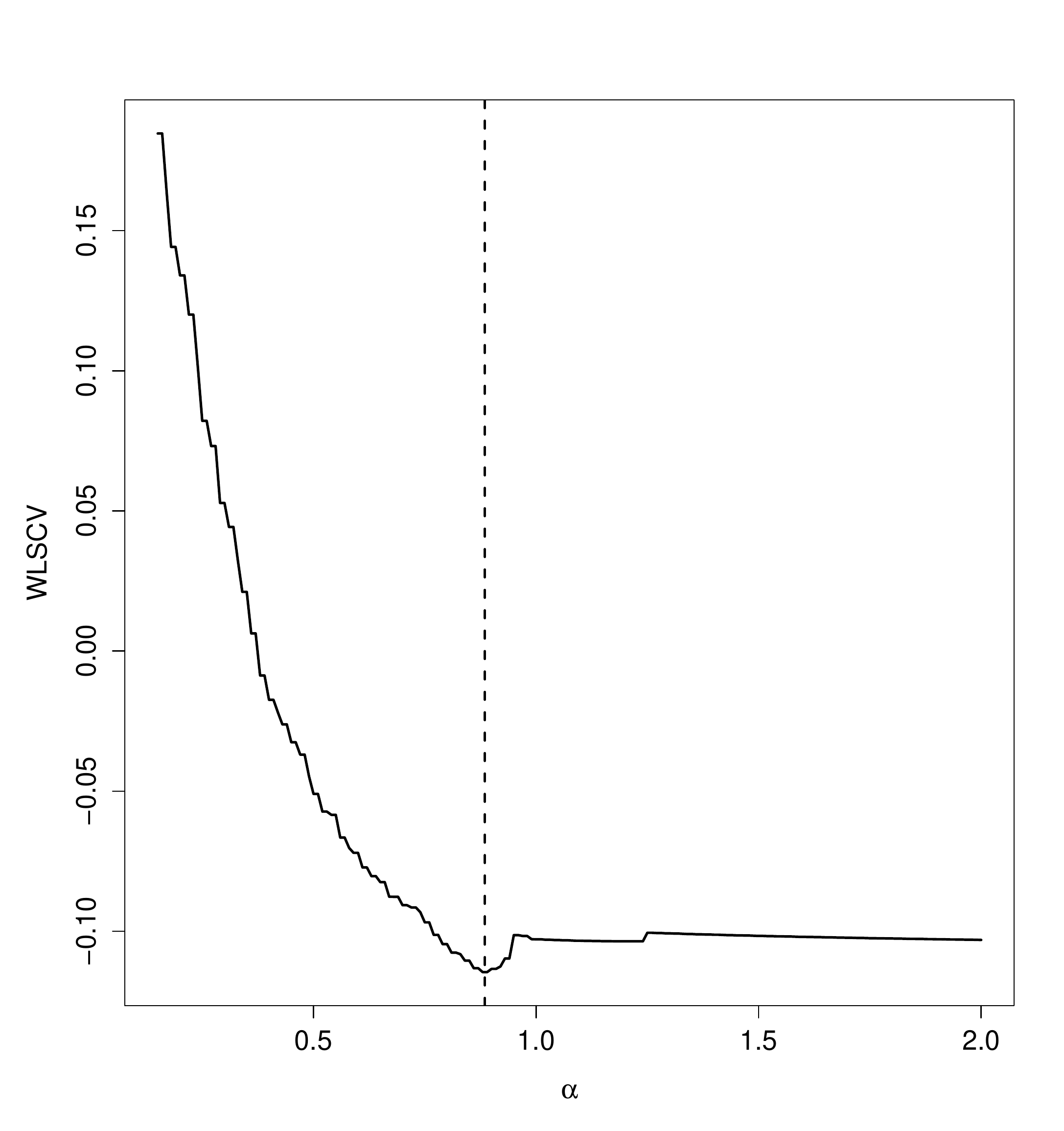}
\caption{Appearance of the Weighted Least Squares Cross-Validation criterion WLSCV2, as a function of $\alpha$, for the the white students enrollment data. The selected value is $\alpha=0.885$. }
\label{fig:WSLCV}
\end{figure}

\ppn Finally the improved probit-transformation kernel density estimator, based on local log-quadratic estimation of $f_S$ (i.e.\ $\tilde{f}_X^{(T2)}$) and $k$-NN bandwidth, was used to estimate $f_X$. The bandwidth was selected via weighted least-squares cross-validation with weight $\omega(s)=\phi(s)/f_S(s)$ (WLSCV2), given the small sample size ($n=56$). The appearance of the Weighted Least Squares Cross-Validation criterion, as in (\ref{eqn:hWLSCV}), is shown as a function of $\alpha$ in Figure \ref{fig:WSLCV}. The selected value was $\alpha=0.885$, resulting in the estimate in Figure \ref{fig:realdat12}. It has a smooth and pleasant appearance, without oversmoothing, and clearly evidences the main features of the underlying density. This nice fit was obtained in a totally automated manner. With the WLSCV1 weight function $\omega(s)=\sqrt{\phi(s)/f_S(s)}$, the selected value of $\alpha$ was 0.92, yielding essentially the same estimate which is therefore not shown. As a comparison, the diffusion kernel 
estimator (computed with its automatic bandwidth) is also shown in the figure, in addition to the previous two Beta and Gaussian Copula kernel estimators. The estimate $\fdiff$ fails to evidence the two features of the density. In fact, the diffusion kernel estimator has a propensity for providing estimates whose derivatives are zero at the boundaries, as shown in \cite{Botev10}'s equations (3) and (4) and the related comments. This is clear from Figure \ref{fig:realdat12}. The obtained estimate can also be compared to Figures 3.9 and 3.17 in \cite{Simonoff96}, which show, respectively, the conventional kernel density estimate and the local log-quadratic estimate directly computed on the raw data,  i.e.\ without transformation. The improvement brought by transforming the data is visually obvious.

\section{Concluding remarks}\label{sec:ccl}

\noindent In this paper, a kernel estimator for unknown densities supported on the unit interval $\Is=[0,1]$ has been devised. The suggested procedure is actually an improved version of the basic idea of the {\it transformation kernel density estimator}, which has been around for a long time but which had never been implemented simply and efficiently. As such, the proposed methodology fills a gap in the literature. A fixed, probit transformation is used in all situations, and it is argued that an approach based on local likelihood methods mostly cures the shortcomings previous `naive' transformation-based estimators suffer from. A range of such `improved' probit-transformation kernel density estimators, varying in the exact nature of the local likelihood method which is used and in the way the bandwidth is selected, has been studied. Asymptotic expressions for the bias and the variance have been derived for the cases of interest. In particular, if the density to estimate is smooth enough (four continuous 
derivatives), it has been shown that the bias of some versions of the estimator is of order $O(h^4)$ without the need for {\it ad hoc} bias correction methods. This is, of course, a property taken over from the local likelihood density estimation methods which enjoy it at the source. In favorable situations, the bias there may even be $o(h^4)$, under the same smoothness assumption. A comprehensive simulation study (23 estimators were studied on 16 different densities on $[0,1]$) has confirmed the very good behavior of the suggested probit-transformation estimators. In particular, an estimator based in local log-quadratic estimation of the density in the transformed domain, coupled with a bandwidth of type $k$-nearest neighbor selected via weighted cross-validation, has positioned itself as the best choice. Remarkably, the cross-validation criterion provides, here, a reliable way of selecting the smoothing parameter, unlike what has sometimes been observed in other settings. On the average, the improved 
probit-transformation estimators have been seen to outperform their main competitors, namely the Beta kernel estimator \citep{Chen99}, the Gaussian Copula estimator \citep{Jones07a,Jones07b} and the diffusion estimator \citep{Botev10}. 

\ppn Some theoretical refinement may be needed to fully contemplate the potential of the suggested method. In particular, local likelihood estimators are known to admit different asymptotics whether the smoothing parameter is forced or not to tend to 0 as $n \to \infty$. The areas of application of the two types of results is evidently quite difficult to delimit in practice, and this is even more the case when the bandwidth is variable and random, like a bandwidth of type $k$-nearest neighbor. As this type of smoothing parameter has been seen to do better than the fixed bandwidth versions, it would be of great interest to derive an appropriate unified asymptotic theory for that case. In this work, only heuristic arguments have been proposed to motivate using $k$-NN bandwidths. 

\ppn On a more applied perspective, one can wonder how to use the present estimator of densities on $[0,1]$ for estimating densities supported on $[0,+\infty)$. \cite{Jones07b} suggested a procedure based on a second transformation: if $Y$ is has a density supported on $\R^+$, then $X = Y/(Y+1)$ (for instance) is supported on $[0,1]$, and the density of $X$ can be estimated with one of the probit-transformation estimator discussed in this paper. This estimated density of $X$ can then be back-transformed to $\R^+$ to provide an estimate of the density of $Y$. This procedure is indeed totally in the same spirit as what has been discussed here, and it would be interesting to study how well it works for curing boundary bias of kernel estimation of densities of positive random variables.

\section*{Acknowledgments}

The author was supported by a Faculty Research Grant from the Faculty of Science, University of New South Wales, Sydney (Australia). The author is also grateful to Dr Z.I.\ Botev (UNSW) for providing the R code implementing the diffusion kernel estimator.




%

\end{document}